%% file: main.tex
\documentclass[sigconf,balance=false]{acmart}
\usepackage{popets}

\setcopyright{popets}
\copyrightyear{2024}

\acmYear{}
\acmVolume{}
\acmNumber{2024}
\acmDOI{}
\acmISBN{}
\acmConference{arXiv preprint}
\usepackage{paralist}
\usepackage{fancyvrb}
\usepackage{subcaption}
\usepackage[flushleft]{threeparttable}
\usepackage{xurl}
\usepackage{csquotes}
\usepackage{setspace}
\usepackage{titlesec}

\titleformat{\paragraph}[runin]{\normalfont\normalsize\bfseries}{\theparagraph}{}{}[.]
\titlespacing{\paragraph}{0em}{0pt}{5pt}

\newcommand{\ie}[0]{\textit{i.e.,}}
\newcommand{\eg}[0]{\textit{e.g.,}}
\newcommand{\etc}[0]{\textit{etc.}}

\newcommand{\vs}[0]{\textit{vs.}}
\newcommand{\wrt}[0]{\textit{w.r.t.}}

\newcommand{\labelname}[1]{{\sffamily#1}}

\theoremstyle{remark}

\makeatletter
\def\thm@space@setup{%
  \thm@preskip=1.5pt plus 1pt
  \thm@postskip=\thm@preskip
}
\makeatother
\theoremstyle{remark}\newtheorem{norm}{Norm}

\begin{document}
\title{Understanding Privacy Norms through Web Forms}

\author{Hao Cui}
\orcid{0000-0002-7574-2004}
\affiliation{%
  \institution{University of California, Irvine}
  \city{} %
  \state{} %
  \country{} %
}
\email{cuih7@uci.edu}

\author{Rahmadi Trimananda}
\orcid{0000-0002-9900-7506}
\affiliation{%
  \institution{University of California, Irvine}
  \city{} %
  \state{} %
  \country{} %
}
\email{rtrimana@uci.edu}

\author{Athina Markopoulou}
\orcid{0000-0003-1803-8675}
\affiliation{%
  \institution{University of California, Irvine}
  \city{} %
  \state{} %
  \country{} %
}
\email{athina@uci.edu}

\input{docs/0_abstract}


\maketitle

\input{docs/1_intro}

\input{docs/2_related-work}
\input{docs/3_crawler}
\input{docs/4_annotation}
\input{docs/5_analyses}

\input{docs/6_privacy_policy}

\input{docs/7_discussion}

\begin{acks}
\label{sec:acknowledgments}
This work was supported in part by the National Science Foundation under award numbers 1956393 and 1900654, and a gift from the Noyce Initiative.
We would like to thank @NyaMisty for the help with the collection of domain categorization data, and Jingning Zhang for her help with the manual validation of classification results.
We appreciate the anonymous PoPETs reviewers for
their insightful feedback that helped improve the paper.

The authors used AI-based tools, including OpenAI ChatGPT~\cite{chatgpt-web} and Grammarly~\cite{grammarly}, to correct typos, grammatical errors, and awkward phrasing throughout the paper.
\end{acks}

\bibliographystyle{ACM-Reference-Format}
\bibliography{reference,online}

\appendix
\input{docs/8_appendices}

\end{document}

%% file: docs/0_abstract.tex
\begin{abstract}

Web forms are one of the primary ways to collect personal information online, yet they are relatively under-studied.
Unlike web tracking, data collection through web forms is explicit and contextualized. Users (i) are asked to input specific personal information types, and (ii) know the specific context (\ie{} on which website and for what purpose). 
For web forms to be trusted by users, they must meet the common sense standards of appropriate data collection practices within a particular context (\ie{} privacy norms).
In this paper, we extract the privacy norms embedded within web forms through a measurement study.
First, we build a specialized crawler to discover web forms on websites. We run it on 11,500 popular websites, and we create a dataset of 293K web forms.
Second, to process data of this scale, we develop a cost-efficient way to annotate web forms with {\em form types} and {\em personal information types}, using text classifiers trained with assistance of large language models (LLMs).
Third, by analyzing the annotated dataset, we reveal common patterns of data collection practices. We find that (i) these patterns are explained by functional necessities and legal obligations, thus reflecting privacy norms, and that (ii) deviations from the observed norms often signal unnecessary data collection.
In addition, we analyze the privacy policies that accompany web forms. We show that, despite their wide adoption and use, there is a disconnect between privacy policy disclosures and the observed privacy norms. 

\end{abstract}

%% file: docs/1_intro.tex
\section{Introduction}
\label{sec:introduction}

Collection of personal information (PI) has been widely studied in the privacy community. %
On websites, PI collection is performed  explicitly through web forms, and/or implicitly through web tracking.
While web tracking has received  much attention~\cite{acar2013fpdetective,englehardt2016online,iqbal2021fingerprinting}, web forms are rarely discussed from a privacy perspective despite being so ubiquitous and designed specifically to collect user inputs.

\begin{figure}
     \centering
     \setlength{\fboxsep}{0pt}
     \begin{subfigure}[b]{0.55\columnwidth}
         \centering
         \fbox{\includegraphics[width=\textwidth]{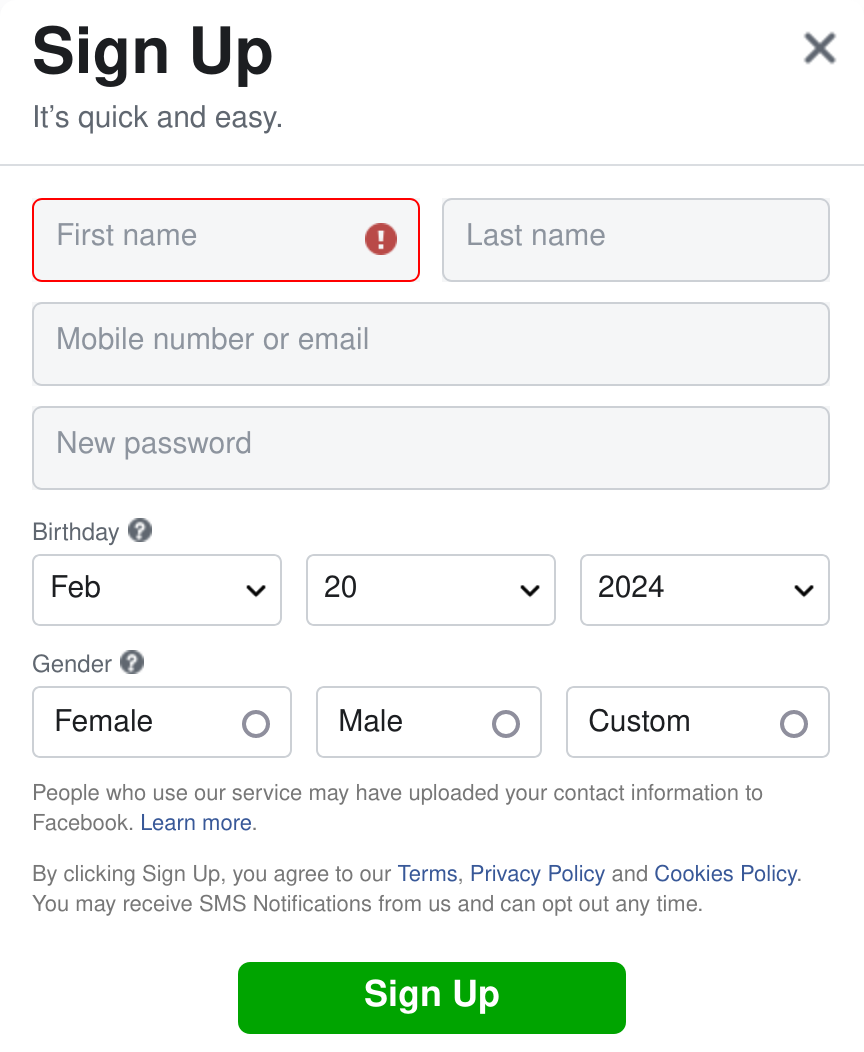}}
     \end{subfigure}
     \hfill
     \begin{subfigure}[b]{0.433\columnwidth}
         \centering
         \fbox{\includegraphics[width=\textwidth]{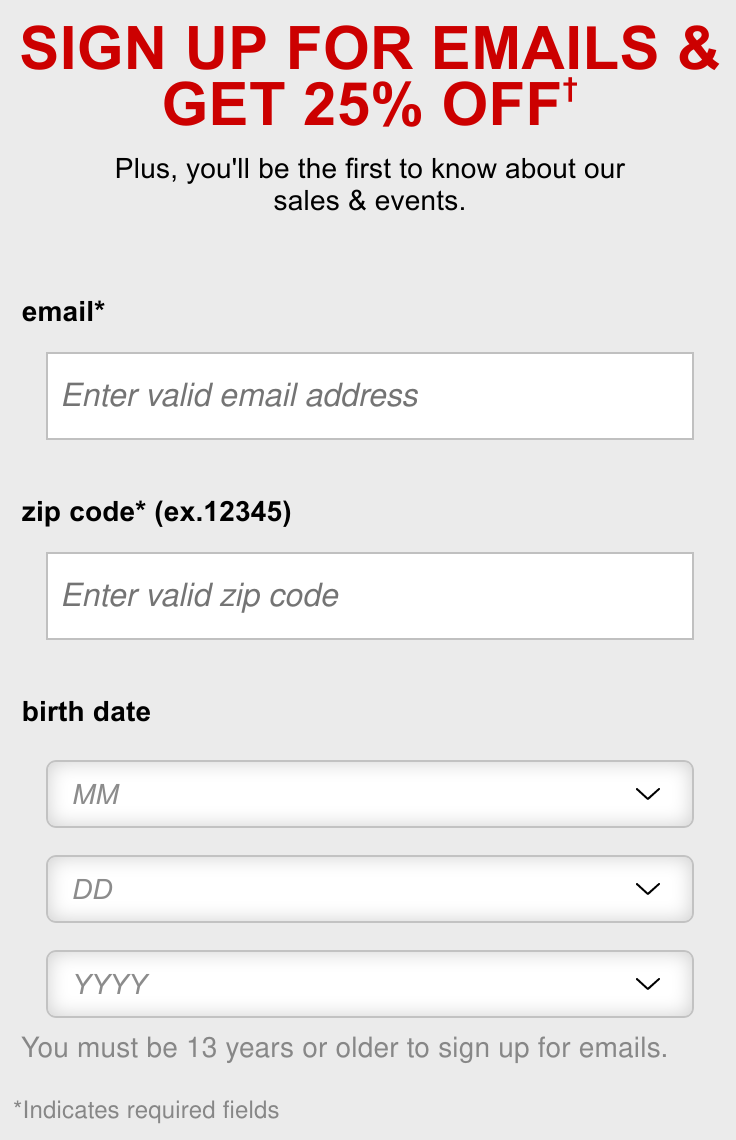}}
     \end{subfigure}
    \caption{
    Web form examples.
    Left: the account registration form on \textit{facebook.com} asks for name, phone number, email address, birth date, and gender.
    Right: the email list subscription form on \textit{macys.com} asks for email address, zip code, and birth date.
    Users have to fill in the requested PI in the fields in order to use the functionality provided by the forms.
    }
    \label{fig:web-form-example}
\end{figure}

PI collection through web forms has unique characteristics that make it explicit and clear.
First, it requires direct user involvement to input PI types that cannot be otherwise automatically collected.
Second, web forms are set up to collect data in specific \textit{contexts}, that is, on specific websites and for specific purposes. %
Figure~\ref{fig:web-form-example} shows two web forms that collect different sets of PI in different contexts.

\paragraph{Privacy Norms}
For web forms to be trusted by users, we argue that they must meet the standards of what is widely considered appropriate PI collection, which we broadly refer to as \textit{privacy norms}. 
For example, account registration on a financial website may require personal tax ID; in contrast, subscribing to a news media site is unlikely to  require such information. 
Privacy norms are often implicit and the research community has explored the topic primarily through user surveys~\cite{shvartzshnaider2016learning,martin2016measuring,apthorpe2019evaluating,abdi2021privacy, apthorpe2018discovering,hoyle2020privacy}.
In this work, we provide a novel approach to understanding privacy norms, by observing common PI collection practices on web forms. Through a measurement study of web forms on popular websites, we connect PI types to contexts%
\footnote{In this work, the \textit{context} of PI collection on a web form refers to the \textit{website category} (\eg{} financial) and the \textit{functionality} of the web form (which reflects the purposes of PI collection, \eg{} account registration).}
in which they are collected.

\paragraph{Web Forms Collection}
To perform measurement at scale, we build a browser-based crawler that discovers and downloads web forms from websites. The crawler is able to simulate clicks on web elements to trigger web forms, including dynamic forms that are created at runtime.
It also implements heuristics to detect links and buttons that likely point to web forms, thus improving efficiency.
We run the crawler on websites from the Tranco list~\cite{tranco} and we collect nearly a million web forms from over 10K English websites.

\paragraph{Dataset Annotation}
We process the HTML code of crawled web forms to infer the functionality (which we refer to as \textit{form type}) and extract \textit{PI types}.
In order to efficiently annotate the huge amount of data, we build a machine learning system that distills the large language model (LLM) into a task-specific text classifier~\cite{tan2024large}.
We use GPT-3.5 Turbo~\cite{openai-gpt}, the state-of-the-art LLM, for unsupervised data labeling, and transfer the knowledge to a smaller form type classifier.
We adopt active learning, with LLM in loop, to improve model generalization on minority labels~\cite{tan2024large,calpric}.
Our methodology is more efficient in terms of monetary and labor costs for labeling than both LLM-only zero-shot classification~\cite{wei2021finetuned,kojima2022large} and traditional manual approach of training data labeling.
After annotation and cleaning up, we create the first annotated web form dataset with 293K web forms from 11,500 websites.

\paragraph{Web Form Analysis}
We analyze the annotated web forms and reveal common patterns of PI collection.
By comparing the collection rates of each PI type across different website categories and form types, we reveal what PI types are collected often in what  contexts. These patterns reflect privacy norms that can be explained by functionality, legal obligations and other reasons, as shown in the following examples from our findings.
(1) The ubiquitous collection of email addresses reflects the perceived non-sensitivity of this PI type.
(2) Other contact information, namely phone numbers and addresses, are used more often by websites that are directly related to real-world services, such as health and finance.
(3) The collection of date of birth and age for account registration can be attributed to children's privacy regulations.
(4) In the financial and health contexts, strict identity verification requirements are evident by the collection of extensive PI types.
Conversely, uncommon PI collection practices that do not align with the privacy norms can indicate excessive PI collection.
The privacy norms identified in our dataset can be used as a baseline to assess data minimization.

\paragraph{Privacy Policy Analysis}
We also compare our observations on web forms to privacy policies -- the legal documents that are supposed to disclose PI collection practices.
We download the privacy policies that accompany web forms.
We find that, while over 90\% of websites provide privacy policies, less than half of the web forms include links within them, indicating that many policies may not be contextualized to explain the web forms.
This is further confirmed by the differences between privacy policy disclosures and observed privacy norms.
We use a state-of-the-art privacy policy analyzer, PoliGraph-er~\cite{cui2023poligraph}, to extract disclosures of PI collection. 
By comparing each website's actual PI collection practices with its privacy policy, we reveal the gap between the two.
On the one hand, some websites do not disclose all the PI types collected in the web forms, indicating possible privacy violations.
On the other hand, some websites simply use blanket disclosures, claiming to collect many PI types that we did not observe, and are unlikely appropriate in the corresponding contexts.
These findings put in question whether privacy policies actually help in understanding websites' PI collection practices.
 
\paragraph{Contributions}
This paper makes the following contributions.

\begin{compactitem}[$\bullet$]
    \item \textit{Measurement of Web Forms:}
    We perform the first, to the best of our knowledge, large measurement study of PI collection through web forms.
    We create a large annotated dataset of 293K web forms on 11,500 popular English websites, which provides a comprehensive view of PI collection practices across different contexts of form types and website categories.
    \item \textit{Understanding Privacy Norms:}
    We propose a novel approach to extracting privacy norms, by analyzing common PI collection practices on web forms.
    We show that these patterns can be attributed to reasons like functionality and legal obligation, which thus can be called privacy norms.
    We also extend our analysis to privacy policies, revealing their misalignment with the norms.
    \item \textit{Methodological Contributions:}
    To facilitate the measurement study, we build a customized crawler to discover and download web forms.
    We also develop a cost-efficient machine learning system to annotate web forms, using a combination of LLM and active learning to help train task-specific classifiers.
\end{compactitem}

\begin{figure}[t!]
    \centering
    \includegraphics[width=\columnwidth]{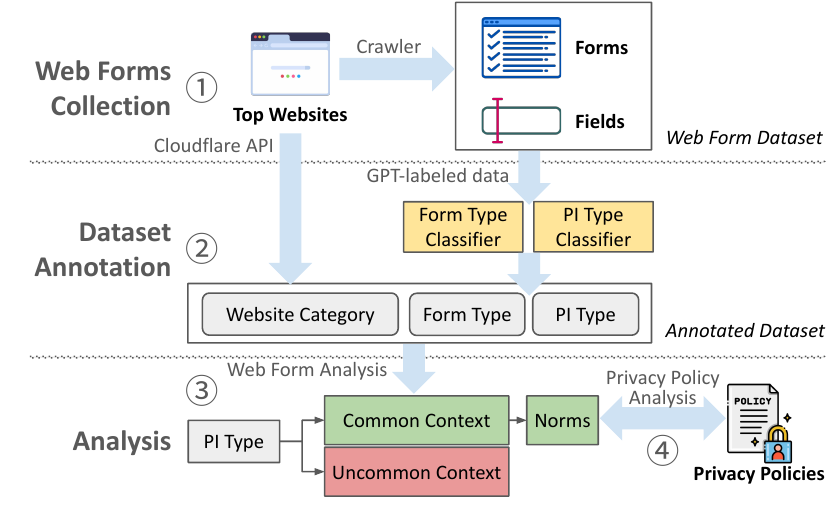}
    \caption{
        Overview.
        \textcircled{1}~We collect web forms from top websites using a customized crawler (Section~\ref{sec:web-forms-collection}).
        \textcircled{2}~We develop a machine learning system to annotate the web forms with form types and PI types (Section~\ref{sec:dataset-annotation}).
        \textcircled{3}~We analyze the web forms to reveal common patterns of PI collection that reflect privacy norms and uncommon cases
        (Section~\ref{sec:analysis}).
        \textcircled{4}~Finally, we also analyze privacy policies to compare the observed norms to disclosed PI collection practices (Section~\ref{sec:privacy-policy-analysis}).
    }
    \label{fig:overall}
\end{figure}

\paragraph{Outline}
Figure~\ref{fig:overall} is an overview of our work.
The rest of the paper is structured as follows.
Section~\ref{sec:related-work} presents related work.
Section~\ref{sec:web-forms-collection} describes the collection of the web form dataset.
Section~\ref{sec:dataset-annotation} describes the dataset annotation methodology.
Section~\ref{sec:analysis} presents the analysis of the web form dataset.
Section~\ref{sec:privacy-policy-analysis} presents the analysis of privacy policies.
Section~\ref{sec:discussion} discusses limitations and future work.

%% file: docs/2_related-work.tex
\section{Background and Related Work}
\label{sec:related-work}

Next, we position our work among related bodies of work on 
 privacy norms, laws, privacy policies and privacy measurement.

\subsection{Contextual Privacy Norms}\label{subsec:privacy-norms}
Privacy norms refer to the common standards of acceptable privacy practices, and specifically in this paper, \textit{acceptable PI collection practices}.
Contextual integrity (CI) is a framework to analyze privacy norms~\cite{nissenbaum2009privacy}. CI states that privacy should be thought as appropriate information flows (rather than just hiding information) that conform to informational (privacy) norms specific to given social contexts.
CI describes such contextual privacy norms using five parameters: (1) data \textit{subject}; (2) \textit{sender} of the data; (3) \textit{recipient} of the data; (4) \textit{information type}; and (5) \textit{transmission principle}, such as the condition or the purposes of PI collection.%
\footnote{For example, it is considered appropriate for a financial institution (\textit{recipient}) to collect name, tax ID and date of birth (\textit{information types}) from clients (\textit{sender / subject}) during account registration for identity verification purposes (\textit{transmission principle}).}

Privacy norms are often implicitly defined by social and cultural norms~\cite{nissenbaum2009privacy}.
A large group of prior work has applied CI to discover these implicit privacy norms. This is usually done through vignette surveys that present participants different contexts (\ie{} combinations of CI parameters) and ask them to score the acceptability~\cite{martin2016measuring}.
\citet{shvartzshnaider2016learning} shows crowdsourcing can be used to discover contextual privacy norms.
Other work surveyed public privacy expectations in specific contexts, such as IoT toy~\cite{apthorpe2019evaluating}, smart home~\cite{abdi2021privacy, apthorpe2018discovering}, online photo sharing~\cite{hoyle2020privacy}.

Our work aims to  understand privacy norms through web forms. Because of their explicit and contextual nature, web forms are governed by implicit privacy norms. The parameters of web forms can be {\em roughly}
 mapped to CI parameters: the website (\textit{recipient}) collects a set of PI (\textit{information types}) from users (\textit{sender \& subject}) for the purposes implied by the form type (\textit{transmission principle}).
We propose a novel way to discover privacy norms by collecting and analyzing web forms from popular websites.
Unlike vignette surveys that present hypothetical situations, the privacy norms we extract are based on actual PI collection practices in the wild. 

\subsection{Laws and Privacy Policy}
\label{sec:law-and-privacy-policy}

Laws and regulations are other sources of privacy norms.
For example, U.S. financial institutions are legally required to collect many PI types from customers for identity verification~\cite{know-your-customer,bank-cip-verification}. In Section~\ref{sec:common-patterns}, we attribute some of the observed privacy norms to legal requirements.
Privacy laws, such as the California Consumer Privacy Act (CCPA)~\cite{ccpa,ccpa-regulations} in the U.S., and the General Data Protection Regulation (GDPR)~\cite{gdpr} in the European Union (EU) and the European Economic Area (EEA), generally do not directly specify what PI types can or cannot be collected.
Instead, they require businesses to provide transparent privacy notices, obtain informed consents (\ie{} {\em notice and consent})~\cite{jordan2022strengths} and/or explain the legal basis for PI collection.
Despite many criticisms about the effectiveness of {\em privacy policies}~\cite{barocas2009notice,okoyomon2019ridiculousness,notice-and-consent-fails}, they have been the main legally-binding mechanism for disclosing PI collection practices.

Privacy policy analysis, including by experts~\cite{wilson2016creation,shvartzshnaider2019going} and via natural language processing (NLP)~\cite{PrivacyGuide,harkous2018polisis,andow2019policylint,bui2021consistency,cui2023poligraph}, has revealed compliance issues in privacy policies.
\citet{shvartzshnaider2019going}, using CI as the analysis framework, report missing contextual details and vague policy language that make the interpretation ambiguous.
Prior work of NLP analysis shows the prevalence of vague, missing or even self-contradicted~\cite{andow2019policylint,okoyomon2019ridiculousness,samarin2023lessons} statements in the privacy policies of mobile apps~\cite{andow2019policylint,andow2020actions,okoyomon2019ridiculousness,lalaine}, virtual reality (VR) games~\cite{trimananda2022ovrseen} and smart home devices~\cite{manandhar2022smart}.
In Section~\ref{sec:privacy-policy-analysis}, we apply PoliGraph-er~\cite{cui2023poligraph}, a state-of-the-art NLP privacy policy analyzer, to analyze privacy policies associated with web forms.
We show the misalignment between privacy policies and observed PI collection, which questions the very role of privacy policies in understanding privacy norms.

Another relevant legal concept is {\em data minimization}.
For example, the CCPA~\cite{ccpa,ccpa-regulations} requires PI collection to be limited to what is
``reasonably necessary and proportionate to achieve the purposes for which the personal information was collected or processed...''
While the law does not define what PI types are necessary under which circumstances, our research, along with prior work on privacy norms~\cite{martin2016measuring,shvartzshnaider2016learning,apthorpe2019evaluating,abdi2021privacy,apthorpe2018discovering,hoyle2020privacy}, provides a baseline for data minimization in different contexts.
In Section~\ref{sec:uncommon-cases}, we discuss how uncommon cases that do not align with privacy norms may indicate \textit{over-collection}, which violates the data minimization principle.

\subsection{Data Collection Measurement}

Extensive research has been conducted to reveal privacy issues of software systems.
In the web ecosystem, web tracking (\ie{} the automatic data collection happening in the background as users interact with websites) has received much attention
~\cite{acar2013fpdetective,englehardt2016online}.
\citet{iqbal2021fingerprinting} report that stateless tracking was used by over a quarter of the top-10K websites as of 2019.
Similar tracking technologies are also used in other non-web platforms, such as smartphones~\cite{andow2020actions,lalaine}, smart TVs~\cite{varmarken2022fingerprintv}, VR headsets~\cite{trimananda2022ovrseen}, and voice assistants~\cite{echoespaper}.

Unlike the opaqueness and invisibility of web tracking, PI collection through web forms appears transparent,%
\footnote{While it is possible that websites with forms also use web tracking, we consider this out of scope, and we focus on measuring the data  explicitly collected by web forms.}
thus better represent privacy norms.
We also note that web forms involve many sensitive PI types that cannot be otherwise automatically collected, such as contact information, government IDs, gender and ethnicity, \etc{}

As for measurements related to web forms, \citet{preibusch2013privacy} find that web users often disclose optional information in survey forms despite the fact that they have a choice.
\citet{lin2020fill,acar2020no} report that malicious web forms can exploit browsers' autofill features to steal credentials.
Prior work also shows that trackers can leak sensitive PI in the web forms to third parties, even before submission~\cite{starov2016you,leaky-forms,chatzimpyrros2019you}.
In contrast to these studies that investigate ``bad'' actors, our work focuses on the ordinary and transparent usage of web forms by first parties to collect PI, and aims to extract privacy norms, instead of anomalies, from them.
To the best of our knowledge, this work is the first paper that analyzes PI collection through web forms in general.

%% file: docs/3_crawler.tex
\section{Web Forms Collection}
\label{sec:web-forms-collection}

In this section, we describe the collection of web form dataset that supports the measurement study.
We develop a browser-based crawler to discover and download web forms.
We run the crawler on over 10K top websites to build the web form dataset.

\subsection{Web Form Crawler}
\label{sec:web-forms-crawler}

The goal of the web form crawler is to discover and save web forms that potentially collect PI. As it is impossible to download the entire website, the main technical challenge is to efficiently discover web pages that contain forms.

We base the crawler on Playwright~\cite{playwright}, a browser automation library.
Playwright allows the crawler to programmatically access websites using a headless but real Google Chrome browser, and simulate actions (\eg{} clicks on elements) in the browser like a real user.
It enables the crawler to interact with dynamic web pages that create web forms using JavaScript code instead of static HTML.

The web form crawler starts by visiting the homepages of websites (\ie{} \texttt{http://<domain>/} or \texttt{http://www.<domain>/}). As the crawler visits web pages, it saves any encountered web forms. It also determines potential next steps (\ie{} what pages to visit next) and adds them to the list of crawler tasks. The task list, which we refer to as  \textit{crawl frontier}~\cite{heydon1999mercator}, is implemented as a priority queue. The crawler assigns each task a priority according to the strategy that we will explain later. %
Tasks with higher priority values are tried first. 
As many web forms need to be triggered dynamically by clicking elements, the crawl frontier does not simply record URLs to visit. Instead, each crawl task is described as a starting URL plus a sequence of click actions on clickable elements (including HTML \texttt{<button>}, \texttt{<a>} elements, and other elements that are bound with JavaScript \texttt{onclick} event handlers).

The main crawler loop is as follows. For each website, the crawl frontier is initialized with a task to access the homepage (and no click actions). In each run, the crawler gets the task with the highest priority from the crawl frontier and runs the following steps:

\begin{compactitem}[$\bullet$]
    \item \textit{Navigation}: The crawler navigates to the starting URL and completes the sequence of click actions specified in the task to restore the full-page state of this crawl task.
    \item \textit{Downloading web forms}: The crawler identifies any web forms (\ie{} HTML \texttt{<form>} elements) on the web page and stores information about each form locally on disk.
    \item \textit{Defining next steps}:
    For each clickable element on the page, the crawler generates a new task as a potential next step. It assigns each task a priority based on the text on it.
    If the element is a hyperlink, the new task will simply set the linked URL as the starting URL. Otherwise, if the element is button-like, the new task will inherit the starting URL and click actions from the current task, and append a new click action to it.
\end{compactitem}

The crawler repeats the steps above to try more and more links and buttons in the crawl frontier, potentially being redirected to new pages in the process, and save any discovered HTML forms.
The crawler limits crawling to each website by skipping tasks that redirect to a different apex domain.

\paragraph{Priority Assignment}
As it is impossible to visit every web page and try every clickable element, we set the crawler to stop after finishing 100 tasks for each website. The crawler thus needs to prioritize steps that likely lead to web forms. To find such steps, the crawler checks how similar the text on each clickable element is to a list of seed phrases. The seed phrases are 100 phrases that are manually curated and indicative of web forms, such as ``Sign Up'', ``Contact Us'', and ``Subscribe''.
The crawler uses \textit{all-MiniLM-L6-v2}, a lightweight sentence transformer model~\cite{reimers-2019-sentence-bert} to compute embedding vectors for both the seed phrases ($t_{s} \in Seeds$) and the text on elements ($t$). Each element is scored as the max cosine similarity from any seed phrases:
$Score(t) = \max_{t_{s} \in Seeds}\allowbreak\mathrm{CosSim}\allowbreak\left(\mathrm{Embed}(t_{s}),\allowbreak\mathrm{Embed}(t)\right)$
where $\mathrm{CosSim}(\mathbf{v_1}, \mathbf{v_2}) = \frac{\mathbf{v_1}\cdot \mathbf{v_2}}{\lVert \mathbf{v_1}\rVert \lVert\mathbf{v_2} \rVert}$.

To avoid crawls of repetitive high-score texts that appear on different pages, the crawler: (1) adds a random number $\varepsilon$ to the score to shuffle steps with similar scores, and (2) discounts the score with a factor that is exponential with the crawl depth ($d$),
assuming that most useful web forms are not nested deeply.
The final priority score is calculated as:%
\footnote{We tested different parameters on a small number of websites. The parameters are chosen empirically to improve the chance of finding more web forms.}
$Priority(t) = 0.9^d (Score(t) + 0.05 \varepsilon),\, \varepsilon \sim U(0, 1)$
Elements with higher priority values are tried first. 

\paragraph{Web Form Discovery and Processing}
The crawler recognizes any HTML \texttt{<form>} elements as web forms. This includes invisible \texttt{<form>} elements which are not yet triggered by the crawler.
The crawler stores the entire HTML code of each web form locally. It also takes advantage of the ability to call Web APIs in the browser to parse the forms. It extracts form fields (\texttt{<input>} elements) and their labels (\texttt{<label>} elements). The label text is a useful feature for classifying PI types (see Section~\ref{sec:data-type-classification}).

The crawler also handles some irregularity of HTML code.
First, many web pages contain irregular web forms that are made of form fields (\texttt{<input>}, \texttt{<select>} and \texttt{<textarea>}) not enclosed by \texttt{<form>} elements. The crawler identifies these isolated field elements and heuristically groups elements that are in close proximity as additional web forms.
Second, some web forms use other text containers that are not \texttt{<label>} elements to show field labels. If the crawler cannot detect the label element of a field, it extracts visible text appearing before the field as the label.

\paragraph{Ethical Considerations}
(1) To avoid interfering with websites' operation, the crawler follows the good practice of limiting the rate per website (at most 10 tasks / minute) and stopping on errors (\eg{} permission denied due to crawler protection).
(2) Although technically possible, the crawler {\em does not} try to fill in any forms or bypass authentication to discover subsequent content.
Due to these (self-)restrictions, the crawler will not discover multi-page forms that appear only after the first page is filled in, or forms that require logging in (\eg{} account settings).

\begin{table}[t!]
\caption{Website category statistics.}
\label{tab:website-category-stats}
\small
\centering
\renewcommand{\arraystretch}{0.7}
\input{tables/website-category-stats}
\end{table}

\subsection{Web Form Dataset}\label{subsec:dataset}

We choose top websites from the Tranco list~\cite{tranco}, a research-oriented domain ranking, as the subjects of our study.%
\footnote{In this paper, a \textit{website} refers to all the content served under an \textit{apex domain}.}
The Tranco list ranks apex domains mainly based on the amount of traffic they receive. Some domains in the list are not meant to be accessed directly (\eg{} content delivery services).
We write a Python script to probe HTTP services on each domain. %
We filter out domains that (1) do not serve HTTP, (2) serve HTTP but redirect to a different domain, or (3) serve non-English content on the homepage.
We also skip websites identified as malware, adult theme or other questionable content by Cloudflare~\cite{cloudflare-family-dns} to prevent our researchers from being exposed to disturbing content when analyzing the data.
We crawled websites in the order listed in the Tranco list, on a server located in California, U.S., from Dec. 2023 through Jan. 2024.
Excluding websites that the crawler did not finish exploring due to errors, we stopped the crawler after collecting approximately 10K valid websites.
In total, the raw web form dataset comprises 938,324 HTML forms from 11,500 websites.%
\footnote{Please see Appendix~\ref{appendix:website-selection} for statistics on discarded and failed websites.
Note that the crawler simply stores all the HTML forms without cleaning. There are many duplicated and irrelevant (\eg{} not collecting PI) samples in the raw dataset.} 
In Section~\ref{sec:processed-dataset}, we describe dataset cleaning that leads to an annotated dataset of 293K forms to be further analyzed.

\paragraph{Website Categorization}
We further categorize websites based on their topics.
The website category is part of the context information that we need to describe the privacy norms (see Sections~\ref{sec:introduction} and~\ref{sec:analysis}).
We use the domain categorization from Cloudflare domain intelligence API~\cite{cloudflare-content-category-api}, which has also been used in prior web measurement studies~\cite{topplingtoplists2022,aworldview2022}.
Cloudflare categorizes websites into a two-level taxonomy, for example, \labelname{Entertainment} (level 1), \labelname{News \& Media} and \labelname{Gaming} (both from level 2).
Table~\ref{tab:website-category-stats} lists the major website categories in our dataset.
Cloudflare may classify each website into zero to multiple categories. In our dataset, 11,263 (97.9\%) websites are in at least one category.

%% file: tables/website-category-stats.tex
\begin{threeparttable}
\begin{tabular}{@{}p{5.9cm}rr@{}}
\toprule
\textbf{Website Category}                               & \multicolumn{2}{l}{\textbf{\# Websites}} \\ \midrule
Technology                                              & 3,206               & 27.9\%              \\
\hspace{2mm} - Technology                                 & 1,983               & 17.2\%              \\
\hspace{2mm} - Information Technology                     & 904                & 7.9\%               \\
Entertainment                                           & 3,041               & 26.4\%              \\
\hspace{2mm} - News \& Media *                            & 992                & 8.6\%               \\
\hspace{2mm} - Video Streaming *                          & 548                & 4.8\%               \\
\hspace{2mm} - Gaming *                                   & 407                & 3.5\%               \\
Business \& Economy                                     & 2,211               & 19.2\%              \\
\hspace{2mm} - Business                                   & 1,662               & 14.5\%              \\
\hspace{2mm} - Economy \& Finance *                       & 509                & 4.4\%               \\
Education                                               & 1,519               & 13.2\%              \\
\hspace{2mm} - Education                                  & 916                & 8.0\%               \\
\hspace{2mm} - Educational Institutions *                 & 538                & 4.7\%               \\
Society \& Lifestyle *                                  & 1,098               & 9.5\%               \\
\hspace{2mm} - Fashion                                    & 188                & 1.6\%               \\
\hspace{2mm} - Clothing                                   & 179                & 1.6\%               \\
\hspace{2mm} - Food \& Drink                              & 164                & 1.4\%               \\
Shopping \& Auctions *                                  & 972                & 8.5\%               \\
\hspace{2mm} - Ecommerce                                  & 733                & 6.4\%               \\
\hspace{2mm} - Shopping                                   & 130                & 1.1\%               \\
Government \& Politics *                                & 921                & 8.0\%               \\
\hspace{2mm} - Politics, Advocacy, and Government-Related & 767                & 6.7\%               \\
\hspace{2mm} - Government                                 & 154                & 1.3\%               \\
Internet Communication *                                & 818                & 7.1\%               \\
\hspace{2mm} - Personal Blogs                             & 418                & 3.6\%               \\
\hspace{2mm} - Information Security                       & 176                & 1.5\%               \\
\hspace{2mm} - Forums                                     & 130                & 1.1\%               \\
Health *                                                & 448                & 3.9\%               \\
Travel *                                                & 328                & 2.9\%               \\ \bottomrule
\end{tabular}
\begin{tablenotes}
  \footnotesize\setstretch{0.8}
  \item \textbf{Note:} Due to a lack of space, we only show the most popular website categories.
  \item * The marked categories, which make up 53.2\% of the dataset, are selected for discussion in Section~\ref{sec:common-patterns}.
\end{tablenotes}
\end{threeparttable}

%% file: docs/4_annotation.tex
\begin{figure}[!t]
    \centering
    \includegraphics[width=\columnwidth]{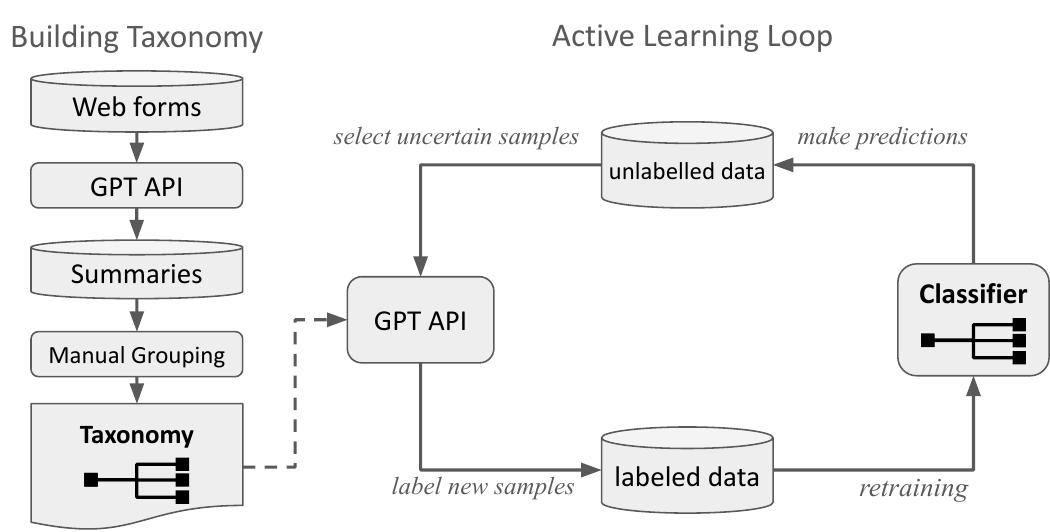}
    \caption{Overview of form type classification.}
    \label{fig:labeling-method}
\end{figure}

\section{Dataset Annotation}
\label{sec:dataset-annotation}

In this section, we describe how we process the raw web form dataset to (1) classify the form type (\ie{} the functionality of the web form), and (2) extract PI types requested in form fields.
To that end, we build machine learning classifiers to efficiently process nearly a million web forms (and many more form fields).

\paragraph{LLM Assisted Classification}
We considered using large language models (LLMs) to annotate the dataset, since modern  LLMs provide a general solution for many NLP tasks, including text classification, without the need to retrain task-specific models~\cite{wei2021finetuned,kojima2022large}.
However, prompting LLMs to process almost 1 million HTML code snippets turns out to be unacceptably slow and costly despite the great accuracy.
Instead, we use LLMs (specifically OpenAI's GPT-3.5 Turbo and GPT-4 Turbo models~\cite{achiam2023gpt,openai-gpt}) to assist classification (for both form type and PI type classification) and to generate training samples (for form type classification) in a  cost-efficient way.

\subsection{Form Type Classification}
\label{subsec:form-classification}

We classify web forms into types according to their functionality. The input for classification is the HTML code of each form saved by the crawler. The functionality is inferred from information in the HTML code, including visible text and HTML attributes. Figure~\ref{fig:labeling-method} shows an overview of our form type classification methodology.

\begin{table}[t!]
\caption{Web form taxonomy for form type classification.}
\label{tab:form-taxonomy}
\resizebox{\columnwidth}{!}{
\centering
\small
\renewcommand{\arraystretch}{2.0}
\linespread{0.5}\selectfont\centering
\input{tables/form-taxonomy}
}
\end{table}

\paragraph{Web Form Taxonomy}
The first step is to build the taxonomy of output labels to use in classification.
To our knowledge, there is no widely accepted taxonomy of web forms.
Table~\ref{tab:form-taxonomy} shows the taxonomy that we create for form type classification.

We derive this taxonomy in a data-driven way. We randomly sample 2,000 web forms%
\footnote{
The taxonomy is intended to cover the functionality of the majority of web forms in our dataset.
In general, the more web forms sampled, the better coverage of the functionality of web forms. However, there are bottlenecks: human workload and budget (the GPT API cost). Also note that returns are diminishing: more samples cover increasingly rare form types that have few samples for training.
For example, Table~\ref{tab:form-taxonomy} shows that, with the 2,000 forms sampled, the rarest form type (Financial Application) was only found on 1.5\% of websites.
}
and prompt the GPT-4 Turbo model to summarize the functionality of each form in a short phrase 
(see Appendix~\ref{appendix:gpt-prompts} for the prompt).
We do not give any specific examples in the prompt to avoid biases. Example responses from GPT-4 are ``newsletter signup'', ``gift purchase'', ``student loan application'', \etc{}
Then, we manually read through the responses and group similar functionality together as one label.
We do not validate the correctness of the GPT-4 responses at this point, but we manually validate the trained classifier later.
In this process, we tried different ways of grouping.
Our objectives are that the labels: (1) cover the functionality of a majority of forms in our dataset; (2) have clear definitions; and (3) have proper granularity so that all labels are sufficiently represented in the dataset.
For the granularity, we use regular expressions to match strings in the web forms to get a rough estimation of the numbers. If a label is too specific to have enough samples for training, we merge it into another label.

The taxonomy in Table~\ref{tab:form-taxonomy} is designed to cover web forms in our dataset.
As such, it does not include all possible web forms on the Internet.
First, due to the limitation of our crawler (see Section~\ref{sec:web-forms-crawler}), we cannot observe forms that require account login, \eg{} account preferences.
Second, it does not include web form types that are not for collecting PI, \eg{} searching forms.
We also note that, to strike a balance between granularity, clarity and coverage, these labels are not fine-grained and by no means accurately describe the diverse use cases of all existing web forms.
The coverage of the taxonomy is illustrated in Table~\ref{tab:form-stats} in Section~\ref{sec:processed-dataset}.

\paragraph{Knowledge Distillation from LLM}
The traditional way of text classification is to train a machine learning classifier from labeled training data. Without the dataset, a modern alternative is zero-shot classification using LLMs~\cite{wei2021finetuned,kojima2022large}.
We only need to input the task description, along with the HTML code to label, and the LLM will do the tedious reading task and give the result.
However, labeling millions of HTML snippets would be costly and slow.
To control the cost, we use OpenAI's GPT 3.5 Turbo model%
\footnote{We use \textit{the GPT} to refer to OpenAI's GPT 3.5 Turbo model for the rest of the paper. The other LLM used in this work is GPT 4 Turbo, which is more powerful than GPT 3.5 Turbo, for taxonomy creation. We use GPT 3.5 Turbo to label web forms for its lower cost (10 times cheaper than GPT 4 Turbo).}
to bootstrap a training dataset, and train a BERT-based text classifier from the GPT-labeled dataset.
This technique is generally known as \textit{knowledge distillation}, which distills the knowledge of a powerful but costly model (\textit{teacher model}) into a smaller task-specific model (\textit{student model})~\cite{tan2024large}.

Specifically, each time we provide the GPT with the HTML code of a web form, along with a prompt that asks it to choose a label in Table~\ref{tab:form-taxonomy} that best describes the form.
We also provide a few more options that are interpreted as a special \labelname{Unknown} label (\ie{} none of the above).
First, we include two common form types that most likely do not collect PI in the prompt: (1) Search forms, for searching contents on websites, and (2) Setting forms, such as cookie consent dialogs, language settings, \etc{}
Second, we prompt the GPT to answer ``Unknown'' if there is not enough information (\eg{} due to incomplete HTML).
See Appendix~\ref{appendix:gpt-prompts} for the full prompt.

In some cases, the functionality of a web form can be ambiguous and more than one category may apply.
To characterize the ambiguity, we take advantage of the GPT's non-determinism to generate \textit{soft labels}.
We set the GPT's temperature parameter to 0.8 (more deterministic than the default 1.0) and get 10 outputs for each sample. The 10 outputs are then converted into soft labels that represent the probabilities of each label in Table~\ref{tab:form-taxonomy}. For example, if a form is used for both account registration and login, the GPT may output the former label 7 times and the latter 3 times, which is converted to soft labels $[0.7, 0.3, 0.0, ...]$.
The \labelname{Unknown} label does not contribute to any of the label probabilities.

\paragraph{Form Type Classifier}
As for the student model, we fine-tuned a MarkupLM model~\cite{li2021markuplm}, a BERT variant, to learn from the GPT-labeled data.
We favor MarkupLM because it is pretrained on HTML code and has a specialized text processor to encode HTML efficiently. Other general-purpose BERT models would have to truncate inputs more often because HTML markups waste a lot of tokens.
A limitation of MarkupLM is that it does not process HTML attributes. In web forms, much information can be found in the attributes. To preserve the information, we improve MarkupLM's processor to extract selected HTML attributes (\eg{} \texttt{placeholder} of \texttt{<input>} elements, and text on buttons).
The classifier is fine-tuned with the GPT-labeled probabilities as the learning objective. To accommodate soft labels, sigmoid activation is used after the classification head, meaning that each label is predicted independently.

\paragraph{Active Learning}
Next, we also need a strategy to choose which samples the GPT should label.
Random sampling is not efficient because our dataset is extremely unbalanced -- in terms of not only label distribution (\eg{} \labelname{Account Login} forms are far more common than \labelname{Role Application} forms) but also other features (\eg{} some websites have more web forms than others).
Imbalanced training data would harm classifier performance on less-represented labels and samples. This is a real problem because we are interested in each form type equally, not just most common ones.
We use \textit{active learning} to interactively expand the training data based on the current classifier performance~\cite{tan2024large,calpric}.
The right part in Figure~\ref{fig:labeling-method} shows the active learning loop. At each round, we run the current classifier to label the dataset. We pick samples for which the classifier is less certain (according to a query strategy that we will explain later), and use the GPT to label them. Then, we retrain the classifier with the expanded training data and go to the next round.

We use a combination of uncertainty-based and random sampling as the active learning query strategy.
First, uncertainty sampling means selecting samples for which the model is least certain. In our case, uncertainty means the predicted probability is close to 0.5 (for any label).
Second, instead of directly picking the probability closest to 0.5, we add some randomness to the target probability and pick the probability closest to $0.5 + \varepsilon$, where $\varepsilon \sim\mathcal N(0, 0.15^2)$. This prevents similar samples from being picked repeatedly.\footnote{
In our dataset, many web forms have the same or similar predicted probabilities of form type labels because
(1) the use of bfloat16 format limits the numeric precision of classifier predictions; (2) there are many web forms that are different in HTML code but have the same embedding after MarkupLM's processor cleaned the code.
The standard deviation of $\varepsilon$ is chosen empirically to sufficiently avoid repetitive samples and not to significantly distort uncertainty-based sampling.
}

The above query strategy helps dataset balance because it picks up uncertain samples of all labels, and ideally less-trained labels tend to have more uncertain samples. However, we still find that the majority labels significantly outnumber the minority ones.
This is a side effect of the soft label -- the classifier is trained to output uncertain probabilities, which voided the assumption that less-trained labels have higher uncertainty.
This is especially an issue in the initial rounds -- minority labels are not predicted at all because of insufficient training.
To workaround the issue, we enforce label balancing by giving the minority labels (\ie{} those that are not predicted) a preference.%
\footnote{
Our objective is to make sure all labels have \textit{enough} training samples.
However, we do not need to enforce perfect balance.
Once the classifier starts to predict all the labels, we stop using the workaround.}
That is, we only consider the probability of those labels when picking up  uncertain samples.

\paragraph{Training \& Validation}
We train the form type classifier as shown in Figure~\ref{fig:labeling-method}.
Specifically, we start by using the GPT to label about 500 samples. To increase the chance of covering the minority labels, we sample web forms that collect different sets of PI types (labeled by the PI type classifier; see Section~\ref{sec:data-type-classification}).%
\footnote{Technically, we make a weighted random selection over all the web forms, with each form weighted by one over the number of forms that collect the same set of PI types.}
In each round, the GPT-labeled data is split into 80\% for training and 20\% for per-epoch validation, and we train the new classifier for 10 epochs.
After model training, we expanded the sample size by about 1,000 using the active learning strategy.
We stopped after 5 rounds, at which point we determined that the classifier performance became stable.
The GPT labeled about 5,400 samples in total.

After we finalize the classifier, we run it to label the entire dataset.
We validate the precision by manually labeling a subset of the dataset.
An author and a volunteer independently labeled 50 random samples for each predicted form type (500 samples in total, no overlap with training data), with a final discussion to resolve any disagreements.
The classifier shows 85.6\% macro average precision.
The full validation results are provided in Table~\ref{tab:form-classification-validation} (left columns).
The precision varies with form types.
\labelname{Account Registration}, \labelname{Account Login}, \labelname{Account Recovery} and \labelname{Payment} forms have high precision, presumably due to their clearly distinguishable functionality and textual features.
For \labelname{Role Application} and \labelname{Financial Application} forms, we find that the classifier can confuse miscellaneous information gathering forms (\eg{} price quote forms for services) with them, resulting in lower precision.
In addition, the functionality of some forms can be ambiguous.
For example, we generally recognize lead generation forms on commercial websites as \labelname{Contact} forms, but similar forms are also used by some websites for collecting user feedback.
In Appendix~\ref{appendix:form-type-classification-examples}, we provide representative examples of correctly classified forms of each type.
In Appendix~\ref{appendix:classification-details}, we show that input sequence length does not significantly affect  precision.

We purposely avoid reporting recall, which measures how many samples of the entire dataset (\ie{} the real distribution) are correctly predicted.
Our validation data is sampled from {\em positive} samples, on which calculating recalls would be misleading, because they do not represent the real distribution.
Due to the extremely unbalanced label distribution (as discussed above and presented in Table~\ref{tab:form-stats} in Section~\ref{sec:processed-dataset}), if we sampled the entire dataset, minority labels would have small support for effective validation.
Furthermore, our dataset is already limited in coverage (due to restrictions explained in Section~\ref{sec:web-forms-crawler}).
Therefore, we conservatively avoid reporting recalls.

\begin{table}[t!]
\centering
\caption{Manual validation of form type and PI classification.}
\small
\renewcommand{\arraystretch}{0.86}
\label{tab:form-classification-validation}
\input{tables/form-classification-validation}
\end{table}

\paragraph{Cost Estimation}
Knowledge distillation and active learning reduce both monetary and labor costs of the classification task.
Here we give a rough cost comparison of our method with pure classifier and pure GPT-based methods.
The traditional way of classifier training with manual dataset creation is labor intensive, namely manual labor is the main cost.
Suppose we were to hire 10 people to do the GPT's task, \ie{} each person labeled 5,400 samples and that labeling each sample took 30 seconds. This would cost, at minimum, 450 human hours, which translates to about 3,260 USD if they were paid based on the current U.S. federal minimum wage~\cite{federal_minimal_wage}.
For the fully GPT-based annotation, LLM inference is the main cost. We estimate that an average web form's HTML code, plus the prompt, is converted to about 2,740 tokens (after we clean up many useless text features). Our raw dataset contains 938K forms.
Even after cleaning up (see Section~\ref{sec:processed-dataset}), there are still about 293K web forms to label.
With the current price of GPT-3.5 API (0.001 USD / 1K tokens), this would cost about 800 USD to classify the entire dataset.%
\footnote{We run the GPT to get 10 predictions for each form. However, the cost does not increase with the number of outputs requested, presumably due to internal caching.}
In our case, we use the GPT to label only 5,400 training samples in total, which costs about 15 USD.

\subsection{PI Type Classification} 
\label{sec:data-type-classification}

We train another classifier to identify the PI type requested in each web form field.
We choose to manually label a small training dataset, for two reasons:
(1) most PI types are sparse (\eg{} tax IDs are collected on few websites) -- it would be more efficient to search for relevant samples to label;
(2) PI types are usually clearly indicated in the text or HTML attributes, so manual labeling is easy if we display the right features (instead of HTML code) to the annotators.

\paragraph{List of PI Types}
Similar to form type classification, we start by prompting the GPT-4 Turbo model to list all the PI types collected in about 4,000 randomly sampled web forms (see Appendix~\ref{appendix:gpt-prompts} for the prompt), and manually build the list of PI types. To maximize the chance of capturing diverse PI types, the forms are sampled across different website categories.
We consider PI types that are generally recognized as personal.
We refer to CCPA Section 1798.140(v) (definition of ``personal information'') and 1798.140(ae) (definition of ``sensitive personal information'') for example PI types~\cite{ccpa}.
We also use regular expressions to estimate if a PI type has a significant presence to support training.

We examine the list of PI types and consider the following 16 labels for PI type classification:
\labelname{Address},
\labelname{Date of Birth},
\labelname{Email Address},
\labelname{Ethnicity},
\labelname{Gender},
\labelname{Tax ID} (\ie{} social security number or equivalents in other countries),
\labelname{Government ID} (\ie{} passport, driver's license number, voter ID or other national IDs),
\labelname{Coarse Location} (city-level or coarser),
\labelname{Postal Code},
\labelname{Bank Account Number} (including credit card number),
\labelname{Person Name},
\labelname{Phone Number},
\labelname{Online Alias} (username or other online IDs used in specific websites),
\labelname{Age} (including age groups),
\labelname{Immigration Status} (including citizenship, nationality, and residency status)
and \labelname{Military Status} (including veteran status).

This list is by no means exhaustive as we exclude many PI types that are rare or limited to few contexts (\eg{} meal preference, size of one's household). For model training purposes, we used three more labels for other common field types:
\labelname{Password},
\labelname{Business Info} (\eg{} contacts of companies, which we do not consider as PI),
and \labelname{Fingerprints} (hidden fields that contain machine-readable information beyond the scope of this study).
These additional labels are equivalent to \labelname{Unknown} labels as in form type classification.

\paragraph{Feature Extraction and Labeling}
To facilitate manual labeling and the use of pre-trained NLP models, we do not directly use HTML code as input features.
The PI types are usually indicated by text and a few HTML attributes (\eg{} \texttt{name}, \texttt{placeholder}, \texttt{type} and \texttt{aria-} attributes) of the field elements and label elements, so we only extract these values and convert them to YAML format which ensures human readability.
For example, the following is the featurized string of a field that requests date of birth information:

\begin{center}
{
\footnotesize
\nopagebreak
\begin{Verbatim}[samepage=true]
  tagName: INPUT
  label: DATE OF BIRTH
  attributes:
  - placeholder: MM/DD/YYYY
  - id: dateOfBirth
\end{Verbatim}
}
\end{center}

This format is used both for training data labeling and as the classifier's input format. One of the authors manually labeled about 2,200 web form fields as the training data, including about 80$\sim$100 positive samples for each of 16 PI types and additional samples of \labelname{Unknown} labels.
We use Label Studio~\cite{Label-Studio}, an open-source dataset annotation platform, to facilitate manual labeling.

\paragraph{Training \& Validation} 
We use SetFit to fine-tune a classifier based on \textit{bge-small-en-v1.5}~\cite{bge_embedding}, a pre-trained sentence transformer model. SetFit is a few-shot learning framework that uses contrastive learning to learn the differences between labels effectively from a relatively small dataset like ours~\cite{tunstall2022efficient}.
To evaluate classifier performance, we randomly select 50 positive samples for each PI type and manually validate the precision, following the same procedure as in Section~\ref{subsec:form-classification}.
We obtain 93.5\% macro average precision.
The result is shown in Table~\ref{tab:form-classification-validation} (right columns).
For most PI types, the field labels are clear so the classifier is expected to learn the textual features well, resulting in over 90\% precision.
We find that false positives are often caused by ambiguous label text.
For example, the label \textit{ID number} is likely recognized as \labelname{Government ID}, but in some contexts the same term refers to other account identifiers.

\subsection{Annotated Dataset}
\label{sec:processed-dataset}

\paragraph{Dataset Cleaning}
We run the form type and PI type classifiers on the 938K web forms.
With the annotations ready, the dataset is cleaned and processed as follows.
First, we discard about 27K web forms on non-English web pages (as determined by HTML \texttt{lang} attributes and lingua-py library~\cite{github-lingua-py}) because our NLP models only understand English.
Second, about 587K forms do not collect any recognizable PI types.
These are mostly search forms, cookie consent dialogs and hidden forms used only for programming purposes.
Lastly, as we are only interested in web forms that collect PI, we discard 31K forms that do not require any personal identifiers (\ie{}
\labelname{\small Address},
\labelname{\small Email Address},
\labelname{\small Government ID},
\labelname{\small Bank Account Number},
\labelname{\small Person Name},
\labelname{\small Phone Number},
\labelname{\small Online Alias},
and \labelname{\small Tax ID}).
As per the CCPA~\cite{ccpa}, {\em personal information} is information that relates to an identified or identifiable individual. It is unclear whether an age verification switch, or gender selection dropdown in a product search form, can count.
Note that the information collected in these discarded forms may still constitute PI when combined with PI collected elsewhere (notably, web tracking~\cite{starov2016you,leaky-forms,chatzimpyrros2019you}).
We conservatively choose precision over guesswork.
After cleaning, there are 292,655 web forms from 11,500 websites in the final annotated dataset for analysis.

\paragraph{Dealing with Duplication}
Websites often reuse components, resulting in duplicated web forms in our dataset.%
\footnote{For reference, if we simply deduplicate the HTML code in the raw dataset, there will be 173,606 unique web forms.}
However, two forms can be visually the same but different in HTML (\eg{} due to randomized tokens), or visually different but serving the same purpose (\eg{} two versions of login forms).
Because of such ambiguity, and to be able to trace forms back to where they are located, we do not remove duplicates from the dataset.
Duplication will not affect any analysis in the following sections, where we report the number of unique websites with specific contexts.

\begin{table}[t!]
\small
\caption{Form type and PI type statistics.}
\label{tab:form-stats}
\renewcommand{\arraystretch}{0.93}
\resizebox{\columnwidth}{!}{
\input{tables/form-stats}
}
\end{table}

\paragraph{Annotated Dataset}
\label{subsec:annotated-dataset}
The annotated dataset has 293K web forms, labeled with information about website categories (Section~\ref{subsec:dataset}), form types (Section \ref{subsec:form-classification}), and PI types (Section \ref{sec:data-type-classification}). Table~\ref{tab:form-stats} shows the occurrences of form type and PI type labels, in terms of {\em number of unique websites} where these occur.

More formally, let $\mathbf{W}$ be the set of possible website categories (see Table~\ref{tab:website-category-stats}),
$\mathbf{F}$ be the set of possible form types (see Table~\ref{tab:form-taxonomy}, also including \labelname{Unknown} label),
$\mathbf{T}$ be the set of possible PI types (see Section~\ref{sec:data-type-classification}),
$x_i$ be the $i$-th web form in the dataset.
The annotated dataset can be described as
$\langle d(x_i), \mathbf{w}(x_i), f(x_i), \mathbf{t}(x_i) \rangle$,
where
$d(x_i)$ is (the apex domain of) the website,
$\mathbf{w}(x_i) \subseteq \mathbf{W}$ is the set of website categories to which the website belong%
\footnote{Recall that a website can be classified into multiple categories (see Section~\ref{subsec:dataset}).},
$f(x_i) \in \mathbf{F}$ is the form type label,
$\mathbf{t}(x_i) \subseteq \mathbf{T}$ is the set of PI types collected by $x_i$.

%% file: tables/form-taxonomy.tex
\begin{threeparttable}
\begin{tabular}{@{}p{2.39cm}p{6.43cm}@{}}
\toprule
\textbf{Label}       & \textbf{Description}                                                      \\ \midrule
Account Registration & For creating new online user accounts.                                  \\
Account Login        & For users to log into existing accounts using their credentials. \\
Account Recovery     & For retrieving or resetting forgotten account credentials. \\
Payment         & For financial transactions, such as bill payments, online purchases or donations. \\
Financial Application & {For applying to financial services like credit cards, loans, financial aid, insurance, investment accounts.} \\
Role Application & For applying to positions such as employment, school admissions, or volunteer opportunities. \\
Subscription & For users to sign up for newsletters, mailing lists, or similar channels of periodic updates. \\
Reservation  &  For users to book services, schedule appointments, register for events, or similar. \\
Contact  & For users to send private messages, inquiries, or feedback to the website owner. \\
Content Submission  &  {For submitting user-generated content like comments, reviews, or ratings, intended to be published on the website.} \\
\bottomrule
\end{tabular}
\begin{tablenotes}
    \footnotesize\setstretch{0.8}
    \item \textbf{Note:} Also see Appendix~\ref{appendix:form-type-classification-examples} for representative examples of each form type.
\end{tablenotes}
\end{threeparttable}

%% file: tables/form-classification-validation.tex
\begin{tabular}{@{}rr|rr@{}}
\toprule
\textbf{Form Type} & \multicolumn{1}{l|}{\textbf{Precision}} & \textbf{PI Type} & \textbf{Precision} \\ \midrule
Role Application       & 0.70                  & Email Address          & 0.98 \\
Financial Application  & 0.72                  & Phone Number           & 1.00 \\
Payment                & 0.90                  & Person Name            & 1.00 \\
Reservation            & 0.84                  & Address                & 0.92 \\
Contact                & 0.78                  & Coarse Location        & 0.98 \\
Content Submission     & 0.92                  & Postal Code            & 1.00 \\
Subscription           & 0.86                  & Age                    & 0.90 \\
Account Registration   & 0.94                  & Date of Birth          & 0.90 \\
Account Login          & 0.98                  & Bank Account Num.      & 0.82 \\
Account Recovery       & 0.92                  & Government ID          & 0.76 \\
\textit{macro average} & 0.856                 & Tax ID                 & 0.96 \\
                       &                       & Online Alias           & 0.88 \\
                       & \multicolumn{1}{l|}{} & Ethnicity              & 0.98 \\
                       & \multicolumn{1}{l|}{} & Gender                 & 1.00 \\
                       & \multicolumn{1}{l|}{} & Immigration Status     & 0.96 \\
                       & \multicolumn{1}{l|}{} & Military Status        & 0.92 \\
\textit{}              &                       & \textit{macro average} & 0.935 \\ \bottomrule
\end{tabular}

%% file: tables/form-stats.tex
\begin{threeparttable}
\begin{tabular}{@{}rr@{\hspace{5pt}}r@{\hspace{2pt}}|rr@{\hspace{5pt}}r@{}}
\toprule
\textbf{Form Type} & \multicolumn{2}{l|}{\textbf{\#Websites}} & \textbf{PI Type} & \multicolumn{2}{l}{\textbf{\#Websites}} \\ \midrule
Role App.            & 959        &  8.3\%   & Email Address       & 9,805 & 85.3\% \\
Financial App.       & 175        &  1.5\%   & Phone Number        & 4,704 & 40.9\% \\
Payment              & 1,096      &  9.5\%   & Person Name         & 7,804 & 67.9\% \\
Reservation          & 413        &  3.6\%   & Address             & 2,244 & 19.5\% \\
Contact              & 6,304      & 54.8\%   & Coarse Location     & 3,792 & 33.0\% \\
Content Submission   & 831        &  7.2\%   & Postal Code         & 2,493 & 21.7\% \\
Subscription         & 4,613      & 40.1\%   & Age                 & 249   &  2.2\% \\
Account Reg.         & 5,138      & 44.7\%   & Date of Birth       & 1,324 & 11.5\% \\
Account Login        & 7,159      & 62.3\%   & Bank Account Num.   & 530   &  4.6\% \\
Account Recovery     & 5,054      & 43.9\%   & Government ID       & 155   &  1.3\% \\
Unknown              & 1,748      & 15.2\%   & Tax ID              & 170   &  1.5\% \\
                     &            &          & Online Alias        & 4,244 & 36.9\% \\
                     &            &          & Ethnicity           & 137   &  1.2\% \\
                     &            &          & Gender              & 466   &  4.1\% \\
                     &            &          & Immigration Status  & 280   &  2.4\% \\
                     &            &          & Military Status     & 69    &  0.6\% \\ \bottomrule
\end{tabular}
\begin{tablenotes}
  \footnotesize\setstretch{0.8}
  \item \textbf{Note:} Please note that a website can have more than one web form and collect more than one PI type. Therefore, the percentages do not sum to 100\%.
\end{tablenotes}
\end{threeparttable}

%% file: docs/5_analyses.tex
\section{Web Form Analysis}
\label{sec:analysis}

\begin{figure*}[t!]
    \centering
    \includegraphics[width=\textwidth,height=77mm]{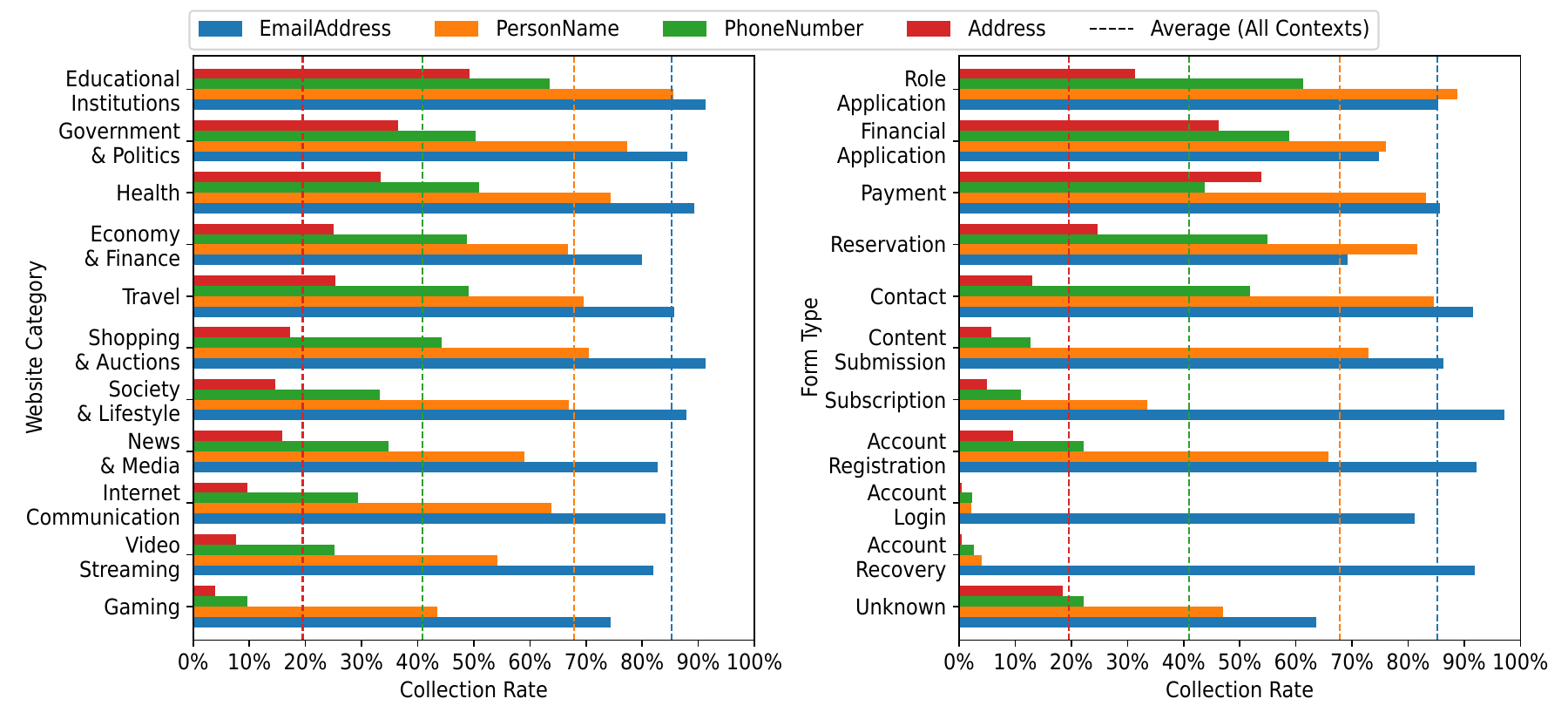}
    \caption{Collection rates of major PI types (contact information) by website category (left) and form type (right).}
    \label{fig:bar_contact}
\end{figure*}

In this section, we analyze patterns of PI collection revealed in the web form dataset.
For each PI type, we seek to understand in what \textit{contexts} (which we define as website category and form type) this PI type is typically collected.
As we argue in Section~\ref{subsec:privacy-norms}, the common
patterns%
\footnote{We use the term \textit{common pattern} to refer to  contexts where a PI type is relatively frequently collected, \ie{} significantly more frequently than average.}
should reflect privacy norms, the common standards on appropriate PI collection.

We mainly compare the \textit{collection rates} of a PI type across contexts.
Formally, we define $N[t | w, f] = \lVert\{ d(x_i) | w {\in} \mathbf{w}(x_i) \wedge f(x_i) {=} f \wedge t {\in} \mathbf{t}(x_i) \}\rVert$, \ie{} the number of \textit{unique websites} in the website category $w$ that use the form type $f$ to collect the PI type $t$.
And specially, we use $p{=}*$, $w{=}*$ or $f{=}*$ to match all labels in that field%
\footnote{That is,
\scalebox{.83}[1.0]{$N[* | w, f]{=}\lVert\{ d(x_i) | w{\in}\mathbf{w}(x_i){\wedge}f(x_i){=}f \}\rVert$},
\scalebox{.83}[1.0]{$N[t | *, f]{=}\lVert\{ d(x_i) | f(x_i){=}f{\wedge}t{\in}\mathbf{t}(x_i) \}\rVert$}, and
\scalebox{.83}[1.0]{$N[t | w, *]{=}\lVert\{ d(x_i) | w{\in}\mathbf{w}(x_i){\wedge}t{\in}\mathbf{t}(x_i) \}\rVert$}.}.
The collection rate of $p$ in the context $(w, f)$ is $P[t | w, f] = \frac{N[t | w, f]}{N[* | w, f]}$.

In this analysis, we compare the collection rate in a specific context, or website category (\ie{} $P[t|w, *]$), or form type (\ie{} $P[t|*, f]$) to the average collection rate over the entire dataset (\ie{} $P[t|*, *]$).

\begin{figure*}[t!]
    \centering
    \includegraphics[width=1.0\textwidth]{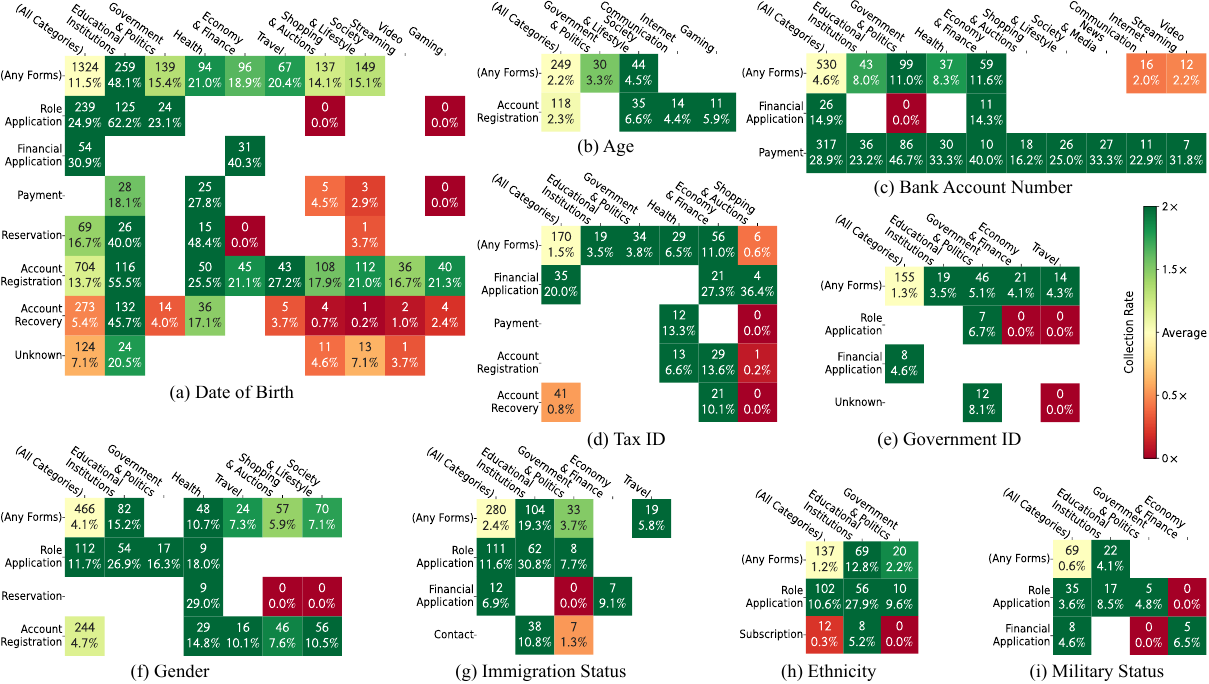}
    \caption{Collection rates of less frequent PI types.
    For each PI type, we show the average collection rate (top-left cell) and contexts where the collection rates are significantly different from the average.
    Each cell shows the collection rate (\ie{} $P[t|w,f]$, the percentage at the bottom) and the number of websites that collect the PI type in the corresponding context (\ie{} $N[t|w,f]$).
    }
    \label{fig:data-types}
\end{figure*}

\paragraph{Results}
We report our web form analysis results in the following figures.
First, Figure~\ref{fig:bar_contact} shows the collection rates of
4 common contact information types:
\labelname{Email Address}, \labelname{Phone Number}, \labelname{Person Name} and \labelname{Address}.
Second, other PI types appear less frequently ($P{<}10\%$) in the dataset. To zoom in into contexts where these PI types are collected frequently, Figure~\ref{fig:data-types} heatmaps show the collection rates by combinations of website categories and form types. Note that some contexts do not have enough samples to be statistically significant.
We use Welch's $t$-test, with $p$ threshold 0.05, to filter out insignificant results.%
\footnote{The collection rate can be viewed as the mean of Bernoulli trials (collect, or not). We compare the collection rate in each context to \textit{other} contexts (\ie{} dataset samples that do not match the current context).
We also acknowledge that the use of $t$-test may lack mathematical rigour because our dataset is subject to labeling errors (see Table~\ref{tab:form-classification-validation}).
}
Due to a lack of space, and to facilitate readability, we only show columns and rows with significantly higher or lower collection rates than average.
We also only choose 11 website categories among others to report (marked in Table~\ref{tab:website-category-stats}).

\subsection{Common Patterns and Privacy Norms}
\label{sec:common-patterns}

Next, we discuss common patterns of web form PI collection, and 
we show that they can be attributed to functional necessities, legal obligations or other norms, thus reflecting privacy norms.
As we collect the dataset in the U.S. and consider only English websites, our discussion focuses on U.S. laws and regulations.

\begin{norm}
\textit{Ubiquity of Email Addresses.}
The most common PI type is \labelname{Email Address}, collected by over 85\% of websites and ubiquitously used across all website categories and form types.
While email address serves as a common contact method and identifier,
some studies have shown that users perceive it as less sensitive  than other identifiers~\cite{markos2017information,SCHOMAKERS2019142,ortlieb2016sensitivity}.
The easiness of creating new email accounts, through either traditional email providers or emerging email masking services~\cite{fastmail-masked-email,firefox-relay}, makes it possible, even for average users, to hide their real identities behind email addresses.

\end{norm}

\begin{norm}
\textit{Phone Numbers and Addresses in Specific Contexts.}
In contrast, Figure~\ref{fig:bar_contact} shows that the collection rates of \labelname{Phone Number} and \labelname{Address} vary significantly across website categories and form types.
Both contact methods have a closer connection to users' real-world identity.
Accordingly, websites with a direct link to real-world services (\eg{} education, government, health, finance, and travel) are more inclined to collect both PI types, reflecting the functional necessities, than online services (\eg{} video streaming, gaming).
\end{norm}

\begin{norm}
\textit{Card Number in Combination with Name and Address for Payments.}
In \labelname{Payment} forms, card numbers (Figure~\ref{fig:data-types}c) are collected along with card holder names and billing addresses (Figure~\ref{fig:bar_contact}).
The norm is tied to standard anti-fraud protocols of the payment card industry, including the PCI DSS standard~\cite{pci_dss} and address verification services~\cite{bank_avs} implemented by major card issuers.
\end{norm}

\begin{norm}
\textit{Age Verification\footnote{Note that our study does not cover all \textit{age-screening mechanisms} in COPPA.
Many websites use standalone age verification forms that only ask for age but no other identifiers. These web forms are not included in the dataset (see Section~\ref{subsec:annotated-dataset}).}
during Account Registration.}
\labelname{Date of Birth} is collected by many website categories for account registration (Figure~\ref{fig:data-types}a). The norm is explained by COPPA requirements. For all \textit{general audience websites}, the COPPA rules require websites to verify the age of users if they collect children's personal information, with some exceptions like one-time contact~\cite{coppa-usc,coppa-rules}.
This requirement is likely agnostic to website categories. 
Account registration by definition collects and stores users' PI and does not qualify for the one-time contact exception.
\end{norm}

\begin{norm}
\textit{Identity Verification in Government, Healthcare and Financial Contexts.}
These services often require extensive PI types, including contact information (Figure~\ref{fig:bar_contact}), government-issued IDs (Figure~\ref{fig:data-types}d and ~\ref{fig:data-types}e) and date of birth (Figure~\ref{fig:data-types}a), for identity verification.
Specifically, healthcare providers are required to verify the identity of a person requesting health information under HIPAA rules~\cite{health-verification-rules}.
Financial institutions are required to verify customer's identity using a combination of name, date of birth, address and government-issued ID numbers, which is known as the ``Know Your Customer'' procedure~\cite{know-your-customer,bank-cip-verification}.
\end{norm}

\begin{norm}
\textit{Personal Attributes in Education and Job Applications.}
\labelname{Gender},
\labelname{Immigration Status},
\labelname{Ethnicity},
and \labelname{Military Status} (Figure~\ref{fig:data-types}f, \ref{fig:data-types}g, \ref{fig:data-types}h and \ref{fig:data-types}i) are more likely collected in the \labelname{Role Application} forms and \labelname{Educational Institutions} contexts. This can be explained by legal requirements.
Specifically, most employers in the U.S. are required to report employment data by ethnicity, gender and national origin~\cite{eeoc-rules}.
And federal contractors are further required to report veteran status data~\cite{veteran-rules}.
The Higher Education Act applies similar requirements to educational institutions~\cite{educational-reporting-requirement}.
\end{norm}

\begin{norm}
\textit{Personal Attributes for Personalization and Marketing.}
Some PI collection patterns are, presumably, linked with personalization and marketing purposes.
\labelname{Gender} and \labelname{Date of Birth} are frequently collected by \labelname{Society \& Lifestyle}, \labelname{Shopping \& Auctions} and \labelname{Travel} categories, especially for account registration (Figure~\ref{fig:data-types}f).
These website categories often provide personalized contents (including ads) based on personal attributes.
We can confirm such usage from the privacy policies of some websites.
For example, \textit{facebook.com}'s privacy policy explicitly states that age and gender information can be used for ``providing ads''~\cite{meta_privacy_policy}.
\textit{casio.com}, the shopping website, mentions that gender can be used for ``marketing and promoting our products'' and ``conducting post-purchase surveys''~\cite{casio_com_privacy_policy}.
\end{norm}

Despite the frequent use, personalization and marketing are not purposes that users find generally worthy or acceptable for giving up their PI~\cite{cit-personalization-survey,apthorpe2018discovering}.
This indicates that privacy norms found in the web forms do not always align with users' expectations.
We hypothesize that web forms represent the websites' view, and not necessarily the users' view, of acceptable PI collection practices.

\subsection{Uncommon Cases}
\label{sec:uncommon-cases}

Our analysis also provides a way to identify rare practices that deviate from the privacy norms.
We scrutinize them to understand their implications, and whether they are appropriate or not.
More specifically, we analyze the data underlying Figures~\ref{fig:bar_contact} and \ref{fig:data-types} to look for contexts with relatively low collection rates.
The first step is to determine ``uncommon'' cases. In this paper, we simply choose a collection rate threshold $p_o$ to find outliers.  Specifically, for web form $x_i$ that collects PI type $t$, we consider such PI collection uncommon if $P[t | w, f_i]{<}p_o$, $\forall w{\in}\mathbf{w}(x_i)$.
For example, with $p_o=2.5\%$ (chosen empirically), there are 855 (7.4\%) websites with uncommon PI collection under this criterion.%
\footnote{The parameter $p_o$, can be tuned by the user of our methodology, and depending on the dataset.
Furthermore, there are other principled methodologies, beyond thresholds, for identifying outliers that can be incorporated.}
Once uncommon cases are identified, the second step is to interpret them, by manually looking into the corresponding forms and websites.
Because our dataset is subject to errors, and our context definition may not fully describe each web form's functionality, we are cautious in interpreting these uncommon cases (see Section~\ref{subsec:limitations} for further discussion).
We find that uncommon cases can indicate unnecessary PI collection or dark patterns, as shown in the representative examples below.

First, \textit{macys.com}, a shopping website, asks for date of birth in its email list subscription form (also shown in Figure~\ref{fig:web-form-example})~\cite{macys_com_email_subscription_form}.
Our dataset shows that
$P[\mathsf{Date\,of\,Birth}\allowbreak|\mathsf{*},\allowbreak\mathsf{Subscription}]=\allowbreak2.3\%$
($\ll 11.5\%$ overall).
The form states that ``You must be 13 years or older'', seemingly indicating the information is for age verification. However, when we actually test the form, we found that the field is optional.
The most relevant explanation in its privacy policy indicates the information is used for ``loyalty program'' (marketing). The misleading language in the form suggests a potential dark pattern to trick users into giving unnecessary information.

Second, \textit{metopera.org}, the website of an opera house, optionally asks for ethnicity for account registration~\cite{metopera_org_account_registration_form}.
Our dataset shows that
$P[\mathsf{Ethnicity}\allowbreak|\mathsf{*},\allowbreak\mathsf{Account\,Registration}]=\allowbreak0.2\%$
($\ll 1.2\%$ overall).
Ironically, its privacy policy states ``We ask that you not send us...  information related to racial or ethnic origin...''~\cite{metopera_org_privacy_policy}, leaving the purpose of PI collection unclear.

Third, some shopping websites ask for users' gender and/or age, linking with the name, in the product review form (\ie{} \labelname{Content Submission}), such as
\textit{colgate.com} and \textit{sleepnumber.com}~\cite{sleepnumber_com_review_form}.
Our dataset shows that
$P[\mathsf{Gender}\allowbreak|\mathsf{*},\allowbreak\mathsf{Content\,Submission}]=\allowbreak1.0\%$
($\ll 4.1\%$ overall) and
$P[\mathsf{Age}\allowbreak|...]=\allowbreak1.8\%$
($< 2.2\%$ overall).
Both privacy policies briefly mention public reviews and merely say that the submitted information may become public~\cite{sleepnumber_com_privacy_policy,colgate_com_privacy_policy}.

Excessive collection violates the data minimization principle in the CCPA~\cite{ccpa,ccpa-regulations} and other privacy laws (see Section~\ref{sec:law-and-privacy-policy}).
While the CCPA does not define what constitutes the minimal set of PI types under which circumstances, our investigation suggests that the observed privacy norms can be used as a baseline to quantify excessive collection.
Also note that, in two of the above cases, the PI types in question are all optional.
In our privacy norm analysis, we do not take optionality into account due to technical challenges in recognizing it.
Separating optional from mandatory PI types when evaluating privacy norms is a useful direction for future work.

%% file: docs/6_privacy_policy.tex
\section{Privacy Policy Analysis}
\label{sec:privacy-policy-analysis}

In this section, we turn our attention to the privacy policies that accompany web forms.
In accordance with the notice requirements of privacy laws (see Section~\ref{sec:law-and-privacy-policy}), one may expect that the privacy policy disclosures cover and, may even conservatively go beyond, what we can observe through web forms, thus helping the understanding of privacy norms. But is this the case?
We extend our analysis by comparing privacy policy disclosures to observed privacy norms.

\subsection{Privacy Policy Availability}
\label{subsec:pp-availability}

We first process the HTML code of web forms and the pages where they locate to search for links (\texttt{<a>} elements) to privacy policies.
We use the same text scoring method as our web form crawler (see Section~\ref{sec:web-forms-crawler}) to fuzzily match link texts and target URLs against seed phrases (\eg{} ``privacy policy'', ``privacy notice'').
Note that we only consider websites that have at least one web form that collects PI%
\footnote{We do not exclude them in Section~\ref{sec:analysis}, in order to keep a consistent number of websites (11,500) and not to overstate the collection rates. However, if a website collect no PI (\eg{} personal blogs), it seems appropriate for it to omit privacy policies.}
(see Section~\ref{sec:processed-dataset}), leaving 10,143 websites for this analysis.
Please see Appendix~\ref{appendix:detection-of-pp} for details on the detection of privacy policy links, including a discussion on the selection of seed phrases.

At the website level, we find privacy policy links on 94.2\% (9,559) of websites, showing the wide adoption of privacy policies.
Table~\ref{tab:web-cats-pp-availability} shows the number of domains without privacy policies by website category.
Notably, \labelname{Educational Institutions} and \labelname{Government \& Politics}, which are more inclined to collect more PI types, have the worst availability.
While missing privacy policies can be a legal compliance issue for businesses, some privacy laws, such as the CCPA, only apply to for-profit entities, which may explain the low availability in these categories.
The same reason may also apply to \labelname{Internet Communication}, which includes subcategories, such as \labelname{Personal Blogs} and \labelname{Forums} (see Table~\ref{tab:website-category-stats}), likely non-profit.

\begin{table}[t!]
\center
\small
\caption{Number of websites without privacy policies by website category.}
\label{tab:web-cats-pp-availability}
\renewcommand{\arraystretch}{0.85}
\input{tables/web-cats-pp-availability}
\end{table}

\begin{table}[t!]
\small
\caption{Number of websites that include privacy policy links inside the form by form type.}
\label{tab:form-type-pp-location}
\renewcommand{\arraystretch}{0.85}
\input{tables/form-type-pp-location}
\end{table}

At the web form level, we further check if privacy policy links are provided \textit{in context} with the forms. Technically, we check if the links are inside \texttt{<form>} elements (see Appendix~\ref{appendix:detection-of-pp} for details). This usually indicates a good practice that allows users to see the notice before submitting the form.
Table~\ref{tab:form-type-pp-location} shows the statistics.
We can see that \labelname{Account Registration} forms are more likely to provide the links close by. \labelname{Account Login} and \labelname{Account Recovery} forms, possibly assuming that the user has read the privacy policies, tend not to provide the links.
Overall, fewer than half of web forms provide the privacy policy links in context.
Many websites put the link in the footer or even on a different web page (usually the homepage).
This raises questions on whether a website-level privacy policy is contextualized to explain the specific PI collection practices happening in the web forms.

\subsection{Content Analysis}
\label{subsec:privacy-policy-content-analysis}

We  look further into the text of privacy policies for  disclosures of collected PI types.
We use PoliGraph-er~\cite{cui2023poligraph}, a state-of-the-art NLP privacy policy analyzer, to extract what PI types are disclosed to be collected.
As PoliGraph-er's PI types do not fully align with ours, we only consider 9 PI types that can be clearly mapped: \labelname{Email Address}, \labelname{Phone Number}, \labelname{Person Name}, \labelname{Address}, \labelname{Date of Birth}, \labelname{Age}, \labelname{Tax ID}, \labelname{Gender}, and \labelname{Ethnicity} for this analysis.
We are able to download privacy policies for 9,013 websites and run PoliGraph-er on them.%
\footnote{Privacy policy links detected on 546 websites are not accessible. Also note that,
as we extract the privacy policy link from each web form and page, there can be multiple privacy policy links on one website. In this case, we combine all the versions by taking the disclosed PI types from all of them into account.}
To avoid false results due to failed HTML parsing, we remove privacy policies that do not disclose any of the 9 PI types,
limiting our analysis to 7,553 websites.
This likely underestimates the number of bad privacy policies (\eg{} due to vague language). However, as PoliGraph-er is optimized for precision rather than recall, we conservatively choose to focus on reliable results.

\begin{figure}[t!]
    \centering
    \includegraphics[width=\columnwidth]{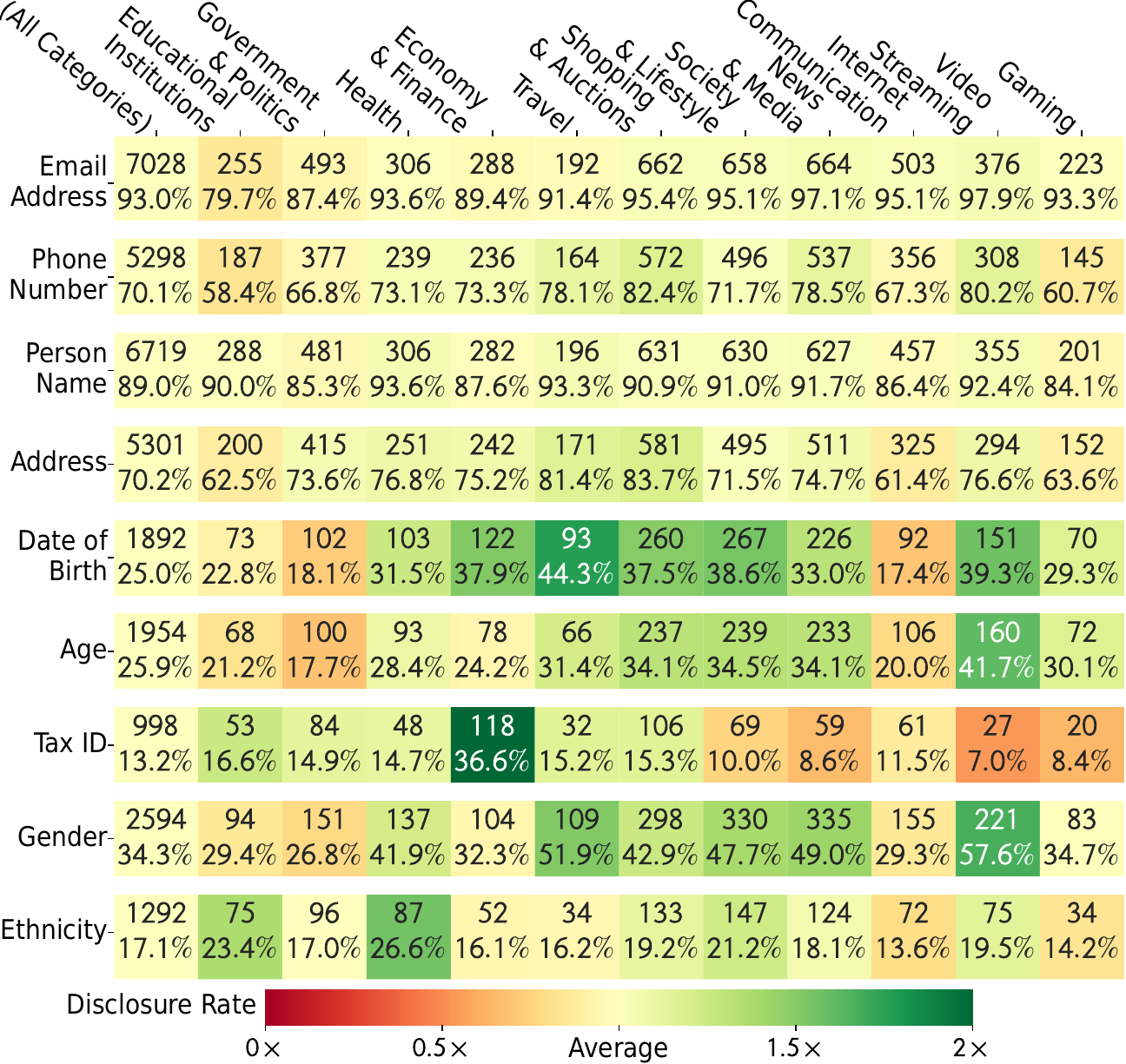}
    \caption{Disclosed PI collection by website category.}
    \label{fig:pp_disclosure_heatmap}
\end{figure}

\paragraph{Patterns of PI Collection Disclosures}
We first check the overall trend of disclosed PI collection, and compare it with privacy norms observed in Section~\ref{sec:common-patterns}.
In Figure~\ref{fig:pp_disclosure_heatmap}, we show, for each PI type, the percentage of websites that disclose the collection of it (\ie{} \textit{disclosure rate}) by website category. It can be seen as collection rates inferred from privacy policies.
We note that the pattern of disclosures is different from our observations in Section~\ref{sec:common-patterns}.

First, \labelname{Educational Institutions} and \labelname{Government \& Policies}, the categories observed to collect extensive PI types in web forms, appear to less often disclose many PI types (\eg{} date of birth, gender, ethnicity) than average.
The contradiction agrees with the previous finding that these categories are doing worse in privacy policies.
For example, \textit{vermont.gov}, the state government's website, provides a privacy policy but only details automatic data collection~\cite{vermont_gov_privacy_policy}. Some web forms on this website require other PI types, including addresses and government-issued ID numbers.
In this case, the collection of these non-disclosed PI types aligns with the observed privacy norms. Indeed, few would be concerned about providing governments with personal identifiers for registering a license~\cite{vermont_gov_educator_registration}.

Then, some other categories, notably \labelname{Video Streaming} and \labelname{News \& Media}, disclose the collection of personal attributes (\eg{} \labelname{Gender}, \labelname{Ethnicity}) and most PI types more frequently than average, contradicting our observations that these online-based services are less PI-intensive.
For example, \textit{netflix.com} claims to collect ethnicity and gender information for ``research surveys''~\cite{netflix_privacy_policy}.
\textit{ft.com}, a news website, claims to collect this information for ``diversity and inclusion goals''~\cite{ft_com_privacy_policy}. 
Both websites provide rather comprehensive privacy policies.
While it is possible our crawler does not capture the corresponding web forms, the takeaway here is that privacy policy disclosures do not correlate well
with observed privacy norms.

\paragraph{Practices vs. Disclosures}

\begin{table}[t!]
\small
\centering
\caption{Privacy policy statements \vs{} PI collection observed.}
\label{tab:pp-content-analysis}
\input{tables/pp-content-analysis}
\end{table}

To further investigate the gap, we compare the disclosures to collected PI types on a per-website basis.
For each PI type, we consider two sets of websites: (1) $C$ are  websites that collect the PI type, as observed in our web form dataset; (2) $P$ are websites that disclose to collect the PI type, as PoliGraph-er determines based on privacy policies. We compare the two and  Table~\ref{tab:pp-content-analysis} summarizes the results.

We distinguish three cases \wrt{} consistency between the two.
(1) $C\&P$ is the ideal case, in which the PI type is both collected and disclosed.
(2) $C\backslash P$ means that the collected PI type is not disclosed, which may indicate privacy violations.
Most prior work focuses on the previous two cases and shows the prevalence of omitted (undisclosed) PI collection~\cite{andow2020actions,trimananda2022ovrseen,cui2023poligraph,lalaine}.
In our case, the high rate of omissions ($P_\mathrm{Omitted}$) for many PI types confirms that. %
(3) However, we are also interested in a third case: $P\backslash C$ means that the PI type is disclosed to be collected, but we do not observe it in our dataset.
The high numbers of both $C\backslash P$ and $P\backslash C$ for many PI types raise the question: \textit{Are privacy policy disclosures really associated with actual PI collection practices?}
In Table~\ref{tab:pp-content-analysis}, we use $\phi$-coefficients as a statistic to measure the association strength between the \textit{observed} PI collection and disclosures.
Interestingly, the association appears to be weak ($<0.20$) for all PI types.

The disconnect is evident in many privacy policies that claim to collect a lot of PI types.
For example, \textit{swarovski.com}, a jewelry brand and a website under \labelname{Society \& Lifestyle} category, literally claims to have collected ``Identifiers such as... social security number; driver’s license number, passport number...'' and the information may even have been disclosed to third parties~\cite{swarovski_privacy_policy}.
In fact, this privacy policy simply copies the definition of identifiers from the CCPA.
\textit{redbox.com}, a video rental company, provides a seemingly detailed privacy policy that explains the purposes for collection and third-party sharing for each PI category~\cite{redbox_com_privacy_policy}. A closer look reveals another case of blanket disclosures -- supposedly collecting  all PI types, each of which can be used for ``other purposes as disclosed'' and shared with ``other service providers''.
On a positive note, it indicates that some PI types are not shared for targeted advertising.
Note that, due to the limited coverage of our web form dataset (see Section~\ref{sec:web-forms-crawler}) and other possible sources of PI collection than web forms, we cannot conclusively verify if these additional PI types ($P\backslash C$ cases) are never collected in the wild.
Nevertheless, ``overly inclusive and broad'' privacy policies have been a common practice for companies to avoid litigation~\cite{nyt-privacy-policy-blog}.

To conclude, our privacy policy analysis hints the gap between (i) the disclosures made in privacy policies and (ii) the observed privacy norms.
This gap is due to both omissions and potentially irrelevant (blanket) information disclosed.
This challenges the very notion that privacy policies help the understanding of websites' PI collection practices.

%% file: tables/web-cats-pp-availability.tex
\begin{tabular}{@{}p{4cm}ll@{}}
\toprule
\textbf{Website Category} & \multicolumn{2}{l}{\textbf{\# w/o privacy policies}} \\ \midrule
Educational Institutions &  53 & (10.4\% of 511) \\
Government \& Politics   &  82 & (10.0\% of 820) \\
Internet Communication   &  56 & (8.00\% of 700) \\
Economy \& Finance       &  23 & (5.07\% of 454) \\
Video Streaming          &  20 & (4.37\% of 458) \\
Gaming                   &  12 & (3.81\% of 315) \\
News \& Media            &  29 & (3.48\% of 834) \\
Travel                   &  11 & (3.75\% of 293) \\
Health                   &  13 & (3.17\% of 410) \\
Society \& Lifestyle     &  23 & (2.62\% of 878) \\
Shopping \& Auctions     &  18 & (1.99\% of 903) \\
{\em All Categories}     & 584 & (5.76\% of 10,143) \\
\bottomrule
\end{tabular}

%% file: tables/form-type-pp-location.tex
\begin{tabular}{@{}p{4cm}ll@{}}
\toprule
\textbf{Form Type} & \multicolumn{2}{l}{\textbf{\begin{tabular}[c]{@{}l@{}}\# w/ privacy policy link\\ in the form\end{tabular}}} \\ \midrule
Account Registration   & 2,309 & (45.0\% of 5,138) \\
Payment                &   401 & (36.6\% of 1,096) \\
Financial Application  &    59 & (33.7\% of 175) \\
Subscription           & 1,491 & (32.3\% of 4,613) \\
Role Application       &   299 & (31.2\% of 959) \\
Contact                & 1,783 & (28.3\% of 6,304) \\
Reservation            &    82 & (19.9\% of 413) \\
Content Submission     &   124 & (14.9\% of 831) \\
Account Login          &   772 & (10.8\% of 7,159) \\
Account Recovery       &   210 & (4.16\% of 5,054) \\ \bottomrule
\end{tabular}

%% file: tables/pp-content-analysis.tex
\begin{threeparttable}
\begin{tabular}{@{}llllll@{}c@{}}
\toprule
\textbf{Data Type} & \textbf{C\&P} & \textbf{C\textbackslash P} & $\mathbf{P_{Omitted}}$ & \textbf{P\textbackslash C} & $\mathbf{P_{NotCollected}}$ & $\phi$ \\ \midrule
Email Address  & 6,859 &  504 & 93.2\% &   169 & 2.40\% & 0.026 \\
Phone Number   & 2,759 &  822 & 77.0\% & 2,539 & 47.9\% & 0.142 \\
Person Name    & 5,287 &  609 & 89.7\% & 1,432 & 21.3\% & 0.043 \\
Address        & 1,282 &  370 & 77.6\% & 4,019 & 75.8\% & 0.085 \\
Date of Birth  &   392 &  580 & 40.3\% & 1,500 & 79.3\% & 0.134 \\
Age            &    71 &  125 & 36.2\% & 1,883 & 96.4\% & 0.039 \\
Tax ID         &    59 &   58 & 50.4\% &   939 & 94.1\% & 0.137 \\
Gender         &   155 &  176 & 46.8\% & 2,439 & 94.0\% & 0.056 \\
Ethnicity      &    29 &   71 & 29.0\% & 1,263 & 97.8\% & 0.037 \\
\bottomrule
\end{tabular}
\begin{tablenotes}
  \footnotesize\setstretch{0.8}
  \item \textbf{Note:} 
$C$ = the set of websites that collect the PI type;
$P$ = the set of websites that disclose the collection of the PI type in their privacy policies; $P_\mathrm{Omitted} = \lVert C\mathrm{\&}P \rVert / \lVert C \rVert$;
$P_\mathrm{NotCollected} = \lVert P\mathrm{\&}C \rVert / \lVert P \rVert$.
\end{tablenotes}
\end{threeparttable}

%% file: docs/7_discussion.tex
\section{Discussion}
\label{sec:discussion}

\subsection{Summary}
In this paper, we propose a novel approach to understanding %
privacy norms through web forms.
We conduct a large-scale measurement study on nearly 293K web forms from 11,500 websites, using a combination of web crawling and NLP classifiers.
Our analysis of the dataset reveals common PI collection patterns that reflect privacy norms, which are influenced by functionality, legal obligations, and other reasons.
We also show that deviations from these norms may indicate excessive data collection.
In addition, our analysis of privacy policies shows a disconnect between the observed norms and the disclosed PI collection practices, thus questioning the role of privacy policies in understanding privacy norms.

\subsection{Limitations}
\label{subsec:limitations}
We acknowledge the following limitations of our study.
\paragraph{Limited and Skewed Coverage}
It is impossible to access every web form on a website. As discussed in Section~\ref{sec:web-forms-crawler}, the crawler does not submit forms or bypass authentication to discover new forms.
Thus, our web form dataset only represents a lower bound, and our analysis may  underestimate the collection rates, which likely affects different PI types unevenly.
For example, in multi-page account registration forms, detailed personal attributes are more likely to be requested after basic contact information is verified, and 
our crawler cannot discover PI types beyond contact information in this case.
To mitigate the issue, we focused on comparing the collection rates of {\em each PI type} across contexts in Section~\ref{sec:analysis}.

Moreover, as discussed in Section~\ref{subsec:dataset}, some websites are excluded for various reasons.
Particularly, we collected data in the U.S. and excluded non-English websites, and our discussion of privacy norms is U.S.-centric.
As explained in Section~\ref{sec:law-and-privacy-policy}, laws and regulations impact privacy norms, so do their regional differences.
For example, in contrast to the CCPA~\cite{ccpa,ccpa-regulations}, which allows opt‐out consent for non-sensitive PI, the GDPR~\cite{gdpr} has stricter consent requirements, mandating opt-in consent, when the consent is the legal basis for processing~\cite{jordan2022strengths}.
Studies have shown regional differences in privacy policy writing~\cite{hosseini2024bilingual} and users' privacy decisions~\cite{cao_android274679}, influenced by legislation and cultural factors.
This being said, our methodology for extracting privacy norms from measurements can be applied beyond English and U.S. websites.

\paragraph{Context Definition}
Recall that we defined the context of PI collection in this study as the combination of website category and form type (functionality).
Due to limitations in categorization and label granularity, the labels may not align with intuition.
For website categories, Cloudflare's categorization is based on content topics~\cite{cloudflare-content-category-api,cloudflare-glossary-content-category}, which do not always indicate the types of services. 
For example, many financial institutions fall in the \labelname{Economy \& Finance} category, but the category also includes financial news websites (which are also in the \labelname{News \& Media} category).
For form types, our taxonomy is coarse and oversimplifies the variety of actual form usages. It also does not directly map to the purposes of PI collection as outlined in privacy policies and laws.
For example, \labelname{Contact} forms cover many lead generation forms (\ie{} marketing purposes) and general customer service forms (\ie{} functional purposes).

\paragraph{Statistical Analyses \vs{} Automated Auditing}
We have focused on statistical analyses to reveal common trends.
In Section~\ref{sec:uncommon-cases}, we investigate whether PI collection that is uncommon compared to the norms can indicate excessive collection.
However, due to the aforementioned limitations, our statistical analysis of privacy norms falls short in the granularity of context definitions.
Our methodology is not meant to be an \textit{automated auditing tool} that reports privacy issues of individual websites.
Therefore, we have been purposely cautious in interpreting uncommon cases and we opted for manual inspection of example cases.
A similar limitation applies to the privacy policy analysis in Section~\ref{subsec:privacy-policy-content-analysis} -- we show the statistical evidence that the observed PI collection
misaligns with privacy policies, but we cannot verify if the (seemingly) overly disclosed PI types are never collected by individual websites.

\paragraph{NLP Methodology}
Our study extensively relies on machine learning models for NLP tasks, which are subject to errors.
For example, despite our effort to improve the form type classifier, it still has low precision for some labels (see Table~\ref{tab:form-classification-validation}).
We also use PoliGraph-er~\cite{cui2023poligraph} for privacy policy analysis.
The tool has a limited recall and may miss some privacy policy disclosures.

\subsection{Implications \& Future Work}

We envision the following applications and future work.

\paragraph{Privacy Risk Assessment}
Our analysis of privacy norms can be used as a baseline to quantify the risk of PI collection.
Section~\ref{sec:uncommon-cases} shows that uncommon cases, which do not align with privacy norms, can indicate excessive PI collection. 
A user-facing tool (\eg{} a web extension) can be implemented based on our infrastructure and analysis to warn users about unusual PI collection, based on form types and website categories.

\paragraph{User Perceptions}
Web forms provide an alternative way to discover privacy norms. It does not replace vignette surveys that directly measure user perceptions.
An interesting research question would be how much the privacy norms observed through web forms align or misalign with users' expectations.
Our high-level intuition is that web forms represent more of websites', rather than user surveys', standpoint \wrt{} the norms.
We cannot refer to prior studies~\cite{shvartzshnaider2016learning,martin2016measuring} to answer the question yet because their surveyed contexts do not fit well with the web form contexts.

\paragraph{Deep Dive into Specific Contexts}
As a reflection on the limitations of our context definition, it would be interesting to perform a deeper analysis of specific contexts.
For example, COPPA imposes strong restrictions on collecting PI from children, thus 
we expect that users' age would impact how web forms collect PI.
For children, some functions might be removed (\eg{} payment forms) or require a restricted set of information (\eg{} in user surveys).
This study would require extensions to our infrastructure, including signing in with different user profiles and collecting web forms on one website for an extended period of time to observe the differences.

To facilitate future extensions and applications, we have released the source code~\cite{webform-code} and the datasets~\cite{webform-artifacts} associated with this study.

%% file: docs/8_appendices.tex
\section{Website Selection Details}
\label{appendix:website-selection}

In Section~\ref{subsec:dataset}, we briefly explain website selection in our dataset, We provide more details in this appendix.

We choose top websites from the Tranco list version 82NJV~\cite{tranco,tranco-82NJV}.
The successfully crawled 11,500 websites (apex domains) rank from 1 (\texttt{google.com}) to 33,515 (\texttt{ebayinc.com}) in the list, \ie{} 22,015 websites in between were not crawled due to various reasons as explained in Section~\ref{subsec:dataset} and detailed below.

The following websites are removed according to the results from the Python HTTP probing script:
\begin{compactitem}[$\bullet$]
    \item 7,924 domains do not resolve, do not serve HTTP(s) on standard ports (TCP 80 and 443), or return HTTP errors (4XX or 5XX) when accessing homepages.
    These include many CDN domains (\eg{} \texttt{googleapis.com} and \texttt{tiktokcdn.com}).
    These also include domains that completely block our script (\eg{} \texttt{whatsapp.com}).
    \item 8,044 websites host non-English contents on their homepages, according to the HTML \texttt{lang} tags (\eg{} \texttt{mail.ru}).
    \item 2,513 websites redirect to different apex domains when accessing their homepages, usually indicating they are not meant to be accessed directly (\eg{} \texttt{amazonaws.com}).
\end{compactitem}

Then, the following websites are filtered out based on the Cloudflare domain intelligence~\cite{cloudflare-content-category-api} results:
\begin{compactitem}[$\bullet$]
    \item 32 are classified as ``public suffixes''. These include some dynamic DNS services that host many websites (\eg{} \texttt{duckdns.org}).
    \item 1,148 websites are classified as unsafe contents (\ie{} any of \labelname{CIPA}, \labelname{Adult Themes}, \labelname{Questionable Content} and \labelname{Blocked} labels).
    \item 255 are classified as ``applications'' that belong to a parent organization -- we only crawl the first among all applications under each organization (\eg{} \texttt{google.co.jp} is removed because \texttt{google.com} supersedes it).
\end{compactitem}

Finally, the crawler failed to finish on 2,099 websites due to errors, including network errors and errors caused by crawler protection.
To rule out sporadic network issues, we retried crawling failed websites at least three times on different days before giving them up.
Top-ranking examples in this category include \texttt{cloudflare.com}, \texttt{icloud.com} and \texttt{msn.com}.

\section{Dataset Annotation Details}
\label{appendix:form-type-classification}

\subsection{GPT Prompts for Dataset Annotation}
\label{appendix:gpt-prompts}

Section~\ref{sec:dataset-annotation} details our dataset annotation methodology, including using the GPT to help build the taxonomy and label training samples that are used to train the form type and PI type classifiers.
We provide the prompts that we used to query the GPT in this appendix.

We use the following prompt for creating the form type taxonomy (\ie{} summarizing web forms) in Section~\ref{subsec:form-classification}:

\begin{displayquote}
\footnotesize\sffamily
Analyze the provided HTML code of a web form, along with the URL and title of the web page to determine the type of the form based on its usage.\\

URL: \texttt{\{...\}}\\
Page Title: \texttt{\{...\}}\\
HTML Code of the Web Form:\\
\verb|```|\\
\texttt{\{...\}}\\
\verb|```|\\

Please use a simple phrase to describe the usage of the form.\\

If insufficient information is available to determine the usage, output "unknown".\\

The response should be in JSON format with a single key "Classification".
\end{displayquote}

We use the following prompt for labeling training samples of form type classification in Section~\ref{subsec:form-classification}:

\begin{displayquote}
\footnotesize\sffamily
Analyze the provided HTML code of a web form, along with the URL and title of the web page to determine the type of the form based on its usage.\\

URL: \texttt{\{...\}}\\
Page Title: \texttt{\{...\}}\\
HTML Code of the Web Form:\\
\verb|```|\\
\texttt{\{...\}}\\
\verb|```|\\

Classify the web form based on its intended usage.\\

Here are the possible categories for classification:\\
- "Account Registration Form": For creating new online user accounts.\\
- \{... {\em see Table~\ref{tab:form-taxonomy} for the full list of label definitions}\}\\
- "Search Form": Used to search or filter website content, typically featuring a search query field and/or filter options.\\
- "Configuration Form": For customizing the user experience on the website, like setting preferences for cookies, language, or display settings.\\

Please choose the category that best describes the form.

If none of the above categories accurately describe the form, suggest a new category.

If the information is insufficient to make a confident classification, label it as "Unknown".\\

Format the response in JSON with one key "Classification".
\end{displayquote}

We use the following prompt for bootstrapping the list of PI types in Section~\ref{sec:data-type-classification}:

\begin{displayquote}
\footnotesize\sffamily
I will provide the HTML code of a web form. Please analyze the form and identify the types of personal data that are being requested in the form fields.\\

"Personal data" (or "personal information") should be understood according to the following definitions in privacy laws:\\

1. **California Consumer Privacy Act (CCPA)**: \{\textit{... quoting CCPA Section 1798.140(v)(1)}\}\\
2. **General Data Protection Regulation (GDPR)**: \{\textit{... quoting GDPR Art. 4 -- Definition of ``personal data''}\}\\

Please analyze the given HTML code of a web form and identify fields that may collect personal data as per these definitions.\\

The output should be in JSON format and include concise and easily interpretable noun phrases that clearly indicate each type of personal data being requested, for example:
\verb|`{|\texttt{"personal\_data\_types": ["Name", "Email Address", "Phone Number"]}\verb|}`|\\

Remember, the focus is on identifying personal data that are being requested in the form. Exclude any data that does not fit abovementioned definitions. If no personal data is being collected, simply output an empty list:\\
\verb|`{|\texttt{"personal\_data\_types": []}\verb|}`|\\

Here is the HTML code of the form:\\
\verb|```|\\
\texttt{\{...\}}\\
\verb|```|
\end{displayquote}

The placeholders (\ie{} \texttt{\{...\}}) are filled in programmatically with corresponding data to be processed.
We also turn on the JSON mode~\cite{openai-json-mode} of the OpenAI GPT API to be able to process the GPT's responses programmatically.

\subsection{Form Type Classification Examples}
\label{appendix:form-type-classification-examples}

\begin{figure*}[t!]
    \centering
    \includegraphics[width=0.99\textwidth]{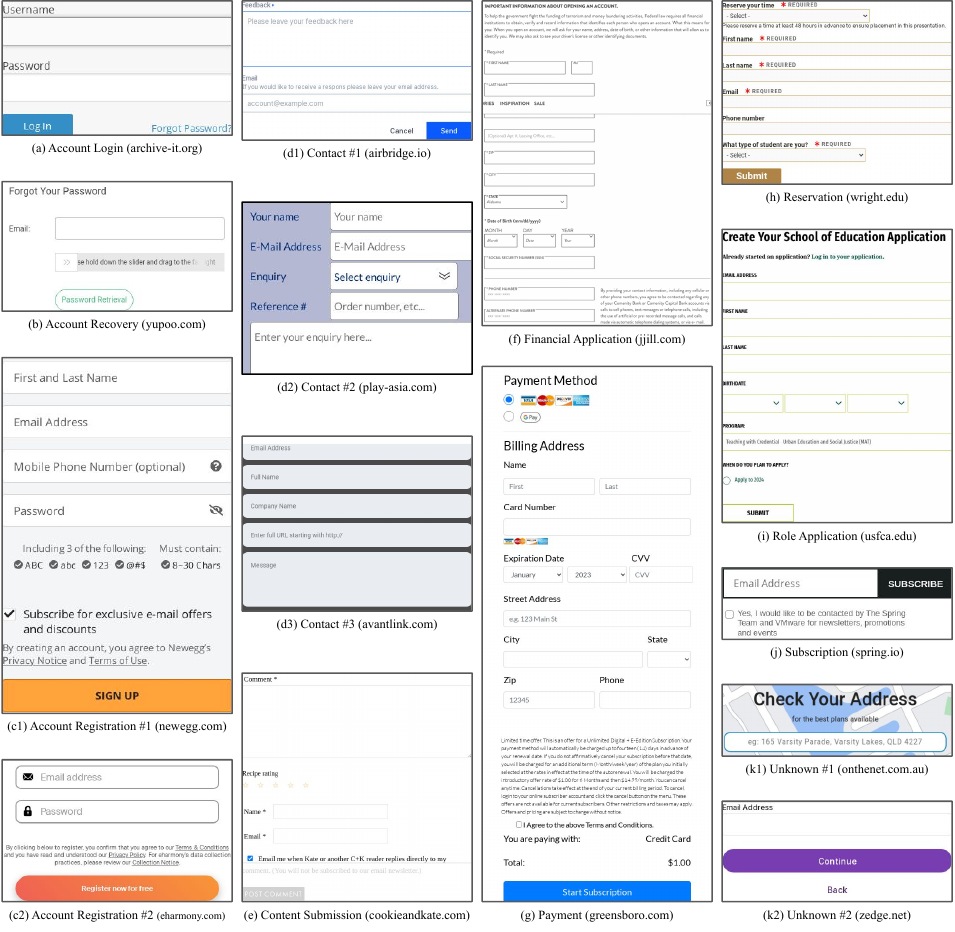}
    \caption{Typical examples of each form type.
    For all form types except \labelname{Unknown}, we pick the examples from our validation data.
    For \labelname{Unknown} forms, we pick the examples from the annotated dataset.}
    \label{fig:form-examples}
\end{figure*}

In Table~\ref{tab:form-taxonomy} in Section~\ref{subsec:form-classification}, we define 10 labels for classifying web forms according to their functionality.
In this appendix, we show representative examples of each form type in Figure~\ref{fig:form-examples}, and summary their visual characteristics below:

\begin{compactitem}[$\bullet$]
    \item {\em Account Login}:
    These forms typically ask for, at least, an account identifier (\eg{} email address or username) and a password (Figure~\ref{fig:form-examples}a).
    The submit button is usually labeled with ``Log In'' or similar phrases.
    \item {\em Account Recovery}: 
    These forms typically ask for a piece of contact information that is also used as the account identifier (Figure~\ref{fig:form-examples}b).
    \item {\em Account Login}: 
    These forms, depending on the services, may ask for various contact information, along with a password to be set (Figure~\ref{fig:form-examples}c1), or a minimal set of information similar to \labelname{Account Login} forms (Figure~\ref{fig:form-examples}c2).
    The submit button is usually labeled with ``Sign Up'', ``Register'' or similar.
    \item {\em Contact}:
    This form type is usually indicated by a free-form text box and requires some contact information.
    The purposes of ``contact'' are broad. Examples include:
    feedback (Figure~\ref{fig:form-examples}d1),
    service inquiry (Figure~\ref{fig:form-examples}d2),
    and lead generation (Figure~\ref{fig:form-examples}d3).
    \item {\em Content Submission}:
    This form type looks similar to \labelname{Contact} forms.
    The difference is the submitted content is intended to be published, which is usually indicated by words ``comment'', ``review'' or similar in the forms (Figure~\ref{fig:form-examples}e).
    \item {\em Financial Application}:
    This form type can require extensive user information, including name, address and tax IDs. The functionality is typically indicated by phrases like ``open an (banking / credit card) account'' in the forms (Figure~\ref{fig:form-examples}f).
    \item {\em Payment}:
    This form type typically asks for payment methods, \ie{} card number, name and address (Figure~\ref{fig:form-examples}g).
    \item {\em Reservation}:
    This form type typically asks for contact information, as well as time preferences for the reservation (Figure~\ref{fig:form-examples}h).
    \item {\em Role Application}:
    This form type can require extensive user information.
    The functionality is typically indicated by phrases like ``(school / job) application'' in the forms (Figure~\ref{fig:form-examples}i).
    \item {\em Subscription}:
    This form type typically asks for very minimum contact information, usually the email address (Figure~\ref{fig:form-examples}j).
    \item {\em Unknown}:
    These forms are not classified into any of the previous types, because
    (1) their functionality does not fit into any form types (\eg{} Figure~\ref{fig:form-examples}k1, the form is used for checking service coverage), or
    (2) there is insufficient information for classification (\eg{} Figure~\ref{fig:form-examples}k2 is the first step of a login form, however, the HTML code and the page title do not clearly indicate the usage).
    The second type can also be counted as false results.
    However, the classifier would have to read additional textual features on the webpages, beyond the HTML code of the web forms themselves, to be able to work.
\end{compactitem}

\subsection{Impact of Input Sequence Length on Form Type Classification}
\label{appendix:classification-details}

In Section~\ref{subsec:form-classification} and Table~\ref{tab:form-stats}, we briefly discuss factors that impact the precision of form type classification.
In response to a question by the reviewers, we additionally analyze the impact of the input sequence lengths on the classifier performance.

In the form type classification, the base MarkupLM model~\cite{li2021markuplm} supports a max input length of 512 tokens, and we truncate any longer HTML inputs.
In this evaluation, we split the 500 validation samples into equal-size quantiles of input sequence lengths (after truncation): $(0, 37]$, $(37, 84]$, $(84, 205]$, $(205, 512]$.
The classifier gets 85.9\%, 84.6\%, 87.1\% and 84.8\% precision, respectively, in the four quantiles.
In addition, on the 55 truncated inputs, the classifier gets 85.5\% precision.
Therefore, we conclude that the input sequence length and input truncation do not significantly affect the classificaiton performance.

\section{Detection of Privacy Policies}
\label{appendix:detection-of-pp}

In Section~\ref{subsec:pp-availability}, we briefly explain how we extract the privacy policy link associated with each website or web form. We provide more details in this appendix.

The web form crawler (see Section~\ref{sec:web-forms-collection}) saves the HTML of all the visited web pages, besides the web forms.
We use a Python script to parse the HTML code to detect the privacy policy links.
It starts by extracting all the links (\ie{} \texttt{<a>} elements) in the HTML code.
Then, similar to the priority assignment strategy explained in Section~\ref{sec:web-forms-crawler}, the text and URL of each link are compared with a list of seed phrases using cosine similarity between the text embeddings.
If the link with the highest cosine similarity score has a score higher than 0.75, it is recognized as the privacy policy link.
Additionally, it also tries to match any link text and URLs that have seed phrases in it (\eg{} \texttt{http://.../privacy-policy-1}).

We use six seed phrases: ``privacy policy'', ``privacy notice'', ``privacy statement'', ``privacy center'', ``privacy \& terms'', and ``privacy \& cookies notice''.
The list was obtained as follows. We started with three commonly used phrases: ``privacy policy'', ``privacy notice'', and ``privacy statement''. Then, we added three more phrases which were later found on websites but have low similarity scores to the initial three, so we could cover more privacy policy links. Our manual validation later shows that the six selected phrases sufficiently cover privacy policy links in our dataset.

To determine if privacy policy links are provided within the forms (\ie{} Table~\ref{tab:form-type-pp-location}), we simply run the detection logic above on the HTML code of individual forms, which effectively limits the detection within the \texttt{<form>} elements.
For each website, we consider all the privacy policy links found on its web pages as its privacy policies (for the availability analysis in Table~\ref{tab:web-cats-pp-availability}, and the content analysis in Section~\ref{subsec:privacy-policy-content-analysis}).

\paragraph{Validation}
We randomly selected 200 websites to manually validate the accuracy of privacy policy link detection, including 17 privacy policy links found inside the web forms, 164 found on the web pages, and 19 without any privacy policy.
One author opened each containing web page, visually checked the form (if the link is detected inside a form), the page footer and also the homepage (in the case of no privacy policy) to search for the privacy policy links.
Overall, the detection results are accurate for 188 (94.0\%) websites.
Of the 12 incorrect results, 4 were due to incorrect third-party privacy policy links being recognized, 6 were due to unexpected link text (\eg{} a website uses ``learn more'' in the cookie consent dialog), and 2 were due to HTML parsing issues.